\pdfoutput=1
\newcommand*{\ATLASLATEXPATH}{latex/}
\documentclass[UKenglish,texlive=2016,cernpreprint,paper=a4]{\ATLASLATEXPATH atlasdoc}

\PreprintIdNumber{CERN-EP-2017-138}

\usepackage{latex/atlaspackage}
\usepackage{latex/atlasbiblatex}
\usepackage[bsm,jetetmiss]{latex/atlasphysics}

\usepackage{multirow}
\usepackage{xspace}

\graphicspath{{logos/}}

\newcommand{\datafifteenlumi}{3.21\xspace}
\newcommand{\datafifteenlumierr}{0.07\xspace} 
\newcommand{\datafifteenlumiwitherr}{\ensuremath{\datafifteenlumi \pm \datafifteenlumierr\,\ifb}\xspace}
\newcommand{\datasixteenlumi}{32.9\xspace}
\newcommand{\datasixteenlumierr}{1.1\xspace} 
\newcommand{\datasixteenlumiwitherr}{\ensuremath{\datasixteenlumi \pm \datasixteenlumierr\,\ifb}\xspace}
\newcommand{\intlumi}{36.1\xspace}

\newcommand{\mt}{\ensuremath{m_\mathrm{T}}\xspace}

\newcommand{\Wjets}{$W$+jets\xspace}
\newcommand{\wjets}{\Wjets}
\newcommand{\Zjets}{$Z$+jets\xspace}

\def\Wln        {$W$ $\rightarrow$ $\ell\nu$ }

\newcommand{\Znn}        {$Z$ $\rightarrow$ $\nu\nu$ }

\newcommand{\ttW}        {\ensuremath{t \bar{t} + W}\xspace}

\newcommand{\ttZ}        {\ensuremath{t \bar{t} + Z}\xspace}
\newcommand{\ttH}        {\ensuremath{t \bar{t} + H}\xspace}

\providecommand{\ttbar}{blah}
\renewcommand{\ttbar}    {\ensuremath{t \bar{t}}\xspace}

\newcommand{\ptvec}      {{\ensuremath{\vec{p}_\mathrm{T}}}}

\newcommand{\dphi}         {\ensuremath{\Delta\phi(j_i,\vetmiss)}}
\newcommand{\njfifty}      {\ensuremath{N_{\mathrm{jet}}^{50}}\xspace}
\newcommand{\njeighty}     {\ensuremath{N_{\mathrm{jet}}^{80}}\xspace}

\newcommand{\vetmiss}{\ensuremath{\vec{E}_{\mathrm T}^\text{miss}}\xspace}
 
\newcommand{\rootgev}{\ensuremath{\,{\textrm{GeV}}^{1/2}}\xspace}
\newcommand{\metsig}{\ensuremath{\met/\sqrt{H_{\mathrm{T}}}}\xspace\xspace}
\newcommand{\METSig}{\ensuremath{\met/\sqrt{H_{\mathrm{T}}}}\xspace\xspace}
\newcommand{\Ht}{\ensuremath{H_{\textrm{T}}}\xspace}

\newcommand{\nbjet}{\ensuremath{{N_\text{$b$-tag}}}\xspace}

\newcommand{\mjsigma}{\ensuremath{M_{\mathrm{J}}^{\Sigma}}\xspace}

\providecommand{\ifb}{\ensuremath{\mathrm{fb}^{-1}}}

\newcommand{\madgraph}{\textsc{MadGraph5}\xspace}

\newcommand{\powhegbox}{\textsc{Powheg-Box}\xspace}
\newcommand{\madspin}{\textsc{MadSpin}\xspace}

\newcommand{\herwigpp}{\textsc{Herwig++}\xspace}
\newcommand{\pythia}{\textsc{Pythia}\xspace}
\newcommand{\evtgen}{\textsc{EvtGen}\xspace}

\newcommand{\amcatnlo}{\textsc{MadGraph5\_aMC@NLO}\xspace}

\newcommand{\sherpa}{\textsc{Sherpa}\xspace}
\newcommand{\comix}{\textsc{Comix}\xspace}
\newcommand{\openloops}{\textsc{OpenLoops}\xspace}
\newcommand{\geant}{\textsc{Geant4}\xspace}

\AtlasTitle{Search for new phenomena with large jet multiplicities and missing transverse momentum using large-radius jets and flavour-tagging at ATLAS in 13 TeV $pp$ collisions}
\author{The ATLAS Collaboration}

\AtlasRefCode{SUSY-2016-13}

\AtlasJournal{JHEP}

\AtlasAbstract{
A search is presented for particles that decay producing a large jet multiplicity and invisible particles.
The event selection applies a veto on the presence of isolated electrons or muons and additional requirements on the number of $b$-tagged jets and the scalar sum of masses of large-radius jets.
Having explored the full ATLAS 2015--2016 dataset of LHC proton--proton collisions at $\sqrt{s}=13\,\TeV$, which corresponds to 36.1~fb$^{-1}$ of integrated luminosity, no evidence is found for physics beyond the Standard Model.
The results are interpreted in the context of simplified models inspired by R-parity-conserving and R-parity-violating supersymmetry, where gluinos are pair-produced.
More generic models within the phenomenological minimal supersymmetric Standard Model are also considered. 
}

\hypersetup{pdftitle={ATLAS document},pdfauthor={The ATLAS Collaboration}}

\begin{document}

\maketitle

\tableofcontents

\section{Introduction}
\label{sec:intro}

As the experiments at the Large Hadron Collider~\cite{Evans:2008zzb} (LHC) continue to amass data from the 13~\TeV\ centre-of-mass energy run, observing the production of heavy resonances remains a principal path in the search for physics beyond the Standard Model (SM), such as supersymmetry (SUSY)~\cite{Golfand:1971iw,Volkov:1973ix,Wess:1974tw,Wess:1974jb,Ferrara:1974pu,Salam:1974ig}.
One distinctive signature of such processes would be an increased incidence of events containing a large number of jets accompanied by missing transverse momentum (the magnitude of which is denoted \met).
These could originate from extended cascade decays of heavy particles through lighter states, which might interact weakly and therefore have remained unobserved due to their low production cross-sections.

A particle spectrum of this nature is exemplified by the pair production of heavy gluinos (\gluino) that decay via long cascade chains, such as through the superpartners of the electroweak and Higgs bosons.
In R-parity-conserving (RPC)~\cite{Farrar:1978xj} SUSY models, these decays culminate in the production of a stable lightest supersymmetric particle (LSP).
Cosmological and other observations prohibit an electrically charged or strongly interacting LSP~\cite{Wolfram:1978gp,Rich:1987jd,Smith:1988ni,Abel:1998wr}, hence the production of these objects, invisible to the detector, would result in missing transverse momentum.
Similarly, large jet multiplicities could also be achieved if the gluinos were to decay via on- or off-shell top squarks (\stopone) or via R-parity-violating (RPV)~\cite{Dreiner:1997uz} couplings.
In the latter case, the LSP could decay within the detector volume, softening the \met spectrum.

Searches by ATLAS were previously reported using smaller quantities of LHC data taken at $\sqrt{s} = 8~\TeV$ from 2011--2012~\cite{SUSY-2011-13,SUSY-2011-19,SUSY-2013-04} and at $\sqrt{s} = 13~\TeV$ in 2015~\cite{SUSY-2015-07}.
Due to the more modest selection on \met, the analysis reported in this paper is sensitive to classes of signals not excluded by related searches performed by ATLAS~\cite{SUSY-2014-06,SUSY-2014-08,SUSY-2015-06,SUSY-2015-08,SUSY-2015-10} and CMS~\cite{CMS-SUS-13-007,CMS-SUS-14-009,CMS-SUS-14-010,CMS-SUS-15-002,CMS-SUS-15-004,CMS-SUS-15-005,CMS-SUS-15-007,CMS-SUS-15-010,CMS-SUS-16-009,CMS-SUS-16-036,CMS-SUS-16-049}.

This paper reports the results of an analysis of \intlumi~\ifb\ of proton--proton ($pp$) collision data recorded at $\sqrt{s} = 13~\TeV$ by the ATLAS experiment~\cite{PERF-2007-01} in 2015 and 2016, scrutinising events that contain significant \met{} and at least seven jets with a large transverse momentum (\pT).
Selected events are further classified based on the presence of jets containing $B$-hadrons ($b$-jets) or on the sum of the masses of large-radius jets.
The $b$-jet selection improves the sensitivity to beyond-the-SM (BSM) signals with enhanced heavy-flavour decays.
Given the unusually high jet multiplicities of the target signatures, large jet masses can originate both from capturing the decay products from boosted heavy particles including top quarks and from accidental combinations~\cite{Cohen:2012yc}.
A key feature of the search is the data-driven method used to estimate the dominant background from multijet production.
Other major background processes include top quark pair production (\ttbar) and $W$ boson production in conjunction with jets (\Wjets).

In the next section the ATLAS detector is described, followed by a description of the accumulated data and simulated event samples in \Sect{\ref{sec:dataMC}}.
Then the event reconstruction and selection are explained in \Sects{\ref{sec:objdef}}{\ref{sec:evtsel}}. The data-driven method to estimate the multijet background and the estimation of systematic errors are in \Sects{\ref{sec:bkgest}}{\ref{sec:stats}}. The result and interpretations are presented in \Sect{\ref{sec:result}} followed by conclusions in \Sect{\ref{sec:conclusion}}.

\section{ATLAS detector}
\label{sec:detector}

\newcommand{\AtlasCoordFootnote}{
ATLAS uses a right-handed coordinate system with its origin at the nominal interaction point (IP) in the centre of the detector and the $z$-axis along the beam pipe.
The $x$-axis points from the IP to the centre of the LHC ring, and the $y$-axis points upwards.
Cylindrical coordinates $(r,\phi)$ are used in the transverse plane, $\phi$ being the azimuthal angle around the $z$-axis.
The pseudorapidity is defined in terms of the polar angle $\theta$ as $\eta = -\ln \tan(\theta/2)$.
Angular distance is measured either in units of $\Delta R_\mathsf{y} \equiv \sqrt{(\Delta \mathsf{y})^{2} + (\Delta\phi)^{2}}$, where $\mathsf{y}$ is the rapidity $1/2 \ln\left((E+p_z)/(E-p_z)\right)$, or in units of $\Delta R$ which is the corresponding quantity in which the pseudorapidity replaces the rapidity.}

The ATLAS detector~\cite{PERF-2007-01} at the LHC covers nearly the entire solid angle\footnote{\AtlasCoordFootnote}\ around the collision point.
It consists of an inner tracking detector surrounded by a thin superconducting solenoid, electromagnetic and hadronic calorimeters,
and a muon spectrometer incorporating three large superconducting toroid magnets.
The inner-detector system (ID) is immersed in a \SI{2}{\tesla} axial magnetic field  and provides charged-particle tracking in the range $|\eta| < 2.5$.

The high-granularity silicon pixel detector covers the vertex region and typically provides four measurements per track.
It is followed by the silicon microstrip tracker which usually provides four two-dimensional measurement points per track.
These silicon detectors are complemented by the transition radiation tracker, which enables the radial extension of tracks with $|\eta| < 2.0$. 
The transition radiation tracker (TRT) also provides electron identification information  based on the fraction of hits (typically 30 in total) above a higher energy-deposit threshold corresponding to the emission of transition radiation.

The calorimeter system covers the pseudorapidity range $|\eta| < 4.9$.
Within the region $|\eta|< 3.2$, electromagnetic calorimetry is provided by barrel and  endcap high-granularity lead/liquid-argon (LAr) electromagnetic calorimeters, with an additional thin LAr presampler covering $|\eta| < 1.8$, to correct for energy loss in material upstream of the calorimeters.
Hadronic calorimetry is provided by the steel/scintillating-tile calorimeter, segmented into three barrel structures within $|\eta| < 1.7$, and two copper/LAr hadronic endcap calorimeters.
The solid angle coverage is completed with forward copper/LAr and tungsten/LAr calorimeter modules optimised for electromagnetic and hadronic measurements respectively.

The muon spectrometer (MS) comprises separate trigger and high-precision tracking chambers measuring the deflection of muons in a magnetic field generated by superconducting air-core toroids.
The precision chamber system covers the region $|\eta| < 2.7$ with three layers of monitored drift tubes, complemented by cathode strip chambers in the forward region, where the background is highest.
The muon trigger system covers the range $|\eta| < 2.4$ with resistive plate chambers in the barrel, and thin gap chambers in the endcap regions.

A two-level trigger system is used to select interesting events~\cite{PERF-2011-02,TRIG-2016-01}.
The Level-1 trigger is implemented in low-latency electronics and uses a subset of detector information to reduce the event rate to below \SI{90}{\kHz}.
This is followed by a software-based High-Level Trigger which reduces the average event rate to about \SI{1}{\kHz}.

\section{Collision data and simulated event samples}
\label{sec:dataMC}

Data recorded by ATLAS during 2015 and 2016 are used in this analysis for background estimation as well as in the final signal region (SR) selections.
Simulated events produced with several Monte Carlo (MC) event generators provide predictions for subdominant background contributions from SM processes producing top quarks and vector bosons.
The main source of background is multijet production, for which predictions are derived directly from data, as described in \Sect{\ref{sec:bkgest:qcdtemplate}}.
Models of potential signals are likewise simulated for analysis optimisation and interpretation of the final results.

\subsection{Data}

Collision events studied for this paper comprise \datafifteenlumiwitherr recorded with good data quality in 2015 with 
a further \datasixteenlumiwitherr recorded during 2016, all with a bunch spacing of 25~ns.
The luminosity uncertainty was derived using beam-separation scans, following a methodology similar to that detailed in Ref.~\cite{DAPR-2013-01}.
Pile-up, i.e.\ additional $pp$ interactions in the same or adjacent bunch crossings, contribute to the signals registered by the detector.
For this dataset, the average number of interactions per bunch crossing ranged up to 52, with a mean of 22.9.

Events were recorded with a variety of triggers. 
During both 2015 and 2016, events were selected by a trigger requiring at least six jets with $\pT>45~\GeV$ and central pseudorapidity, $|\eta|<2.4$. 
In addition, in 2015 events were triggered by requiring the presence of at least five jets with $\pT>70~\GeV$, and in 2016 with a trigger requiring at least five jets with $\pT>65~\GeV$ and $|\eta|<2.4$.

Minimum data quality requirements are imposed to ensure that only events are used in which the entire ATLAS detector was functioning well.
These, for example, exclude data corruption in the ID and calorimeters, excessive noise and spurious jets produced by non-collision backgrounds~\cite{DAPR-2012-01,ATLAS-CONF-2015-029}.

\subsection{Simulated event samples}

All simulated events were overlaid with multiple $pp$ collisions simulated with the soft QCD processes of \pythia 8.186~\cite{Sjostrand:2007gs} using the A2 set of parameters (A2 tune)~\cite{ATL-PHYS-PUB-2012-003} and the MSTW2008LO parton distribution functions (PDFs)~\cite{Martin:2009iq}.
The simulated events were required to pass the trigger, and were weighted such that the pile-up conditions match those of the data.
The response of the detector to particles was modelled with an ATLAS detector simulation~\cite{SOFT-2010-01} based fully on \geant~\cite{Agostinelli:2002hh}, or using fast simulation based on a
parameterisation of the performance of the ATLAS electromagnetic and hadronic calorimeters~\cite{ATL-SOFT-PUB-2014-01} and on \geant elsewhere.

\subsubsection{Background process simulation}

\label{sec:dataMC:background}

For the generation of \ttbar and single top quarks in the $Wt$- and $s$-channels \powhegbox~v2~\cite{Nason:2004rx,Frixione:2007vw,powheg-box,Alioli:2009je,Re:2010bp,Alioli:2011as} was used with the CT10 PDF sets~\cite{CT10pdf} in the matrix element calculations.
Electroweak $t$-channel single-top-quark events were generated with \powhegbox~v1, 
using the four-flavour scheme for the next-to-leading-order (NLO) matrix element calculations, together with the fixed four-flavour PDF set CT10f4~\cite{CT10pdf}, and with top quark decays using \madspin~\cite{madspin}, preserving all spin correlations. 
The parton shower, fragmentation, and the underlying event were simulated using \pythia~v6.428~\cite{Sjostrand:2006za} with the CTEQ6L1 PDF sets~\cite{Pumplin:2002vw} and the Perugia 2012 tune (P2012)~\cite{perugia}.
The top quark mass was set to $172.5$\,\GeV.
The \evtgen~v1.2.0 program~\cite{evtgen} was used to model the properties of the bottom and charm hadron decays. 
Simulated \ttbar{} events were normalised to the cross-section calculated to next-to-next-to-leading order (NNLO) in perturbative QCD, including soft-gluon resummation to next-to-next-to-leading-logarithm (NNLL) order \cite{top++}.

Events containing \ttbar and additional heavy particles -- comprising three-top, four-top, \ttW, \ttZ and $\ttbar + WW$ production -- were simulated at leading order (LO) in the strong coupling constant $\alphas$, using \madgraph~v2.2.2~\cite{Alwall:2014hca} with up to two additional partons in the matrix element, interfaced to the \pythia~8.186 parton shower model.
The A14 set of \pythia 8 parameters was used~\cite{ATL-PHYS-PUB-2014-021}, together with the NNPDF2.3 LO PDF set~\cite{NNPDF}.
The predicted production cross-sections were calculated to NLO as described in Ref.~\cite{Alwall:2014hca}.
In addition, \ttH events were simulated at NLO using \amcatnlo v2.2.3~\cite{Alwall:2014hca}, with the NNPDF3.0 NLO PDF set~\cite{Ball2015} used in the matrix element calculation, and again interfaced to \pythia~8.186 for the parton shower, 
with the A14 tune and the NNPDF2.3 LO PDFs. 

Events containing a $W$ or $Z$ bosons associated with jets were simulated using the \sherpa 2.2.1 generator. Matrix elements were calculated for up to two partons at NLO and four partons at LO using the \comix~\cite{comix} and \openloops~\cite{openloops} matrix element generators and merged with the \sherpa parton shower~\cite{sherpashower} using the ME+PS@NLO prescription~\cite{mepsnlo}.
The NNPDF3.0 NNLO PDF set was used in association with a tuning performed by the \sherpa authors. 

Diboson processes with four charged leptons, three charged leptons + one neutrino, or two charged leptons + two neutrinos, were simulated using \sherpa~v2.1.1~\cite{sherpa2}.
The matrix element calculations contained all diagrams with four electroweak vertices.
They were calculated for up to one (for 4$\ell$, 2$\ell$+2$\nu$) or zero additional partons (for 3$\ell$+1$\nu$) at NLO and up to three additional partons at LO using the \comix and \openloops matrix element generators and merged with the \sherpa parton shower using the ME+PS@NLO prescription. 
The CT10 PDF set was used in conjunction with dedicated parton shower tuning developed by the \sherpa authors.
An identical procedure was followed to simulate diboson production with one hadronically decaying boson accompanied by one charged lepton and one neutrino, two charged leptons or two neutrinos, where the calculations included one additional parton at NLO for $ZZ\to 2\ell+q\bar{q}$ and $ZZ\to 2\nu+q\bar{q}$ only, and up to three additional partons at LO.

Theoretical uncertainties were considered for all these simulated samples. 
By far the most important process simulated in this analysis is \ttbar production, and several samples are compared to estimate the uncertainty in this background.
Samples were produced with the factorisation and renormalisation scales varied coherently, along with variations of the $h_{\mathrm{damp}}$ parameter in \powhegbox and with parameters set for more/less radiation in the parton shower~\cite{ATL-PHYS-PUB-2015-011}.
Additionally, to account for uncertainties from the parton shower modelling and generator choice, the nominal sample was compared to samples generated with \powhegbox and \amcatnlo, interfaced to \herwigpp~\cite{Bahr:2008pv}.
The comparison with samples which vary the amount of additional radiation contributes the largest uncertainty in the signal region predictions. 

Full simulation was used for all background MC samples, ensuring an accurate representation of detector effects.
Further details of samples can be found in Refs.~\cite{ATL-PHYS-PUB-2015-011,ATL-PHYS-PUB-2016-002,ATL-PHYS-PUB-2016-003,ATL-PHYS-PUB-2016-004,ATL-PHYS-PUB-2016-005}.

\subsubsection{Supersymmetric signal models}

\label{sec:dataMC:signal}

A number of supersymmetric signal model samples are simulated to permit the interpretation of the search results in terms of supersymmetric parameters.
Substantial cross-sections are possible for production of gluinos, superpartners of the gluon, whose cascade decays result in a large multiplicity of jets, which may also exhibit an unusually high heavy-flavour content or atypically large masses.

The first is a simplified model, in which gluinos are pair-produced and then decay through an off-shell squark via the cascade:
\begin{align*}
\gluino   & \to q + \bar{q}' + \chinoonepm \hspace*{0.3cm} (q = u,d,s,c), \\
\chinoonepm & \to W^{\pm} + \ninotwo \\
\ninotwo  & \to Z + \ninoone,
\end{align*}
where the quarks are only permitted to be from the first two generations. The parameters of the model are the masses of the gluino, $m_{\gluino}$, and the lightest neutralino, $m_{\ninoone}$.
The mass of the $\chinoonepm$  is constrained to be 
$(m_{\gluino} + m_{\ninoone})/2$,
and the mass of the $\ninotwo$ is set to 
$(m_{\chinoonepm} + m_{\ninoone})/2$.
A diagram of this ``two-step'' simplified model is shown in Figure~\ref{samples:fig:feynman_diagrams}(a).

A second type of SUSY model is drawn from a two-dimensional subspace of the 19-parameter phenomenological minimal supersymmetric Standard Model (pMSSM)~\cite{Djouadi:1998di,Berger:2008cq}, motivated in part by models not previously excluded by the analysis of Ref.~\cite{SUSY-2014-08}. An example 
pMSSM process is shown in Figure~\ref{samples:fig:feynman_diagrams}(b).
These models are selected to have a bino-dominated neutralino \ninoone, kinematically accessible gluinos, and an intermediate-mass Higgsino-dominated multiplet, containing two neutralinos (the \ninotwo\ and \ninothree) and a chargino (the \chinoonepm).
The masses of these particles are varied by changing the SUSY soft-breaking parameters $M_3$ (for the gluino) and $\mu$ (for the Higgsinos), while $M_1$ (for the \ninoone) is held constant at 60 GeV.
In order that other SUSY particles remain kinematically inaccessible, the other parameters, defined in Ref.~\cite{SUSY-2014-08}, are set to 
$m_A=M_2=3~\TeV,$ $A_\tau=0$, $\tan\beta=10$,
$A_t=A_b=m_{(\tilde{e},\tilde{\mu},\tilde{\tau}){\mathrm{L}}}=m_{(\tilde{e},\tilde{\mu},\tilde{\tau}){\mathrm{R}}}=m_{\tilde{q}{\mathrm{L}}(1,2,3)}=m_{(\tilde{u},\tilde{c},\tilde{t}){\mathrm{R}}}=m_{\
(\tilde{d},\tilde{s},\tilde{b}){\mathrm{R}}}=5~\TeV$.
These values ensure theoretically consistent mass spectra, and produce a mass for the lightest scalar Higgs boson close to 125~\GeV.
Mass spectra with consistent electroweak symmetry breaking are generated using \textsc{Softsusy}~3.4.0~\cite{Allanach:2001kg}.
The decay branching ratios are calculated with \textsc{Sdecay/Hdecay}~1.3b/3.4~\cite{Djouadi:2006bz}, and when $m_{\chinoonepm} \lesssim 500$ GeV and $m_{\gluino} \gtrsim 1200$ GeV the predominant decays are 
$\gluino \to t + \bar{t} + \ninotwo (\ninothree)$ and $\gluino  \to t + \bar{b} + \chinoonepm$, 
with \ninotwo\ (\ninothree) decaying to $Z / h + \ninoone$ and \chinoonepm to $W^{\pm} + \ninoone$.
When these decays dominate, they lead to final states with many jets, several of which are \bjets,  but relatively little \met.
This renders this search particularly sensitive compared to most other SUSY searches which tend to require high \met.
At higher $m_{\chinoonepm}$ and lower $m_{\gluino}$, the decay $\gluino \to q q \ninoone$ becomes dominant and this search starts to lose sensitivity.
This model is labelled in the following figures as `pMSSM'.

Gluino-mediated top squark ($\stopone$) production, with the squarks being off-shell, is also a good match for the target final state. 
This scenario is characterised by the pair-production of gluinos followed by their decay with 100\% branching ratio to $\ttbar + \ninoone$, through a virtual top squark. 
Naturalness arguments for supersymmetry favour light gluinos, top squark, and Higgsinos, so this final state is very well motivated. 
 Figure~\ref{samples:fig:feynman_diagrams}(c)  shows a diagram for the off-shell process.

\begin{figure}
\centering
\begin{tabular}{cc}
\subfloat[Two-step decay]{\includegraphics[width=5cm]{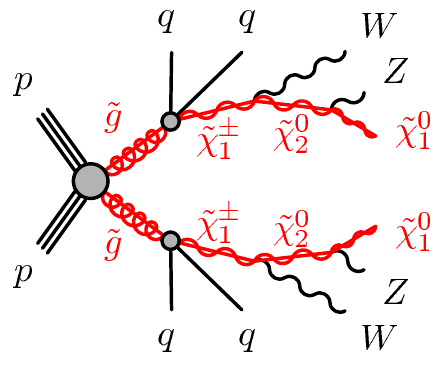}}&
\subfloat[pMSSM]{\includegraphics[width=5cm]{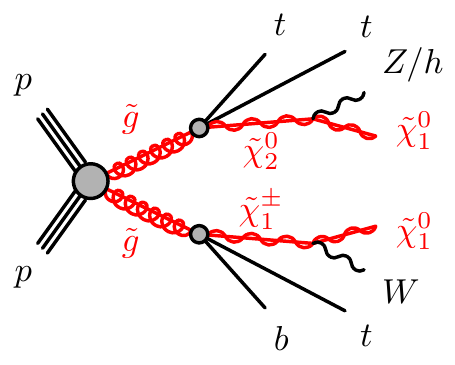}}\\
\subfloat[Off-shell top squarks]{\includegraphics[width=5cm]{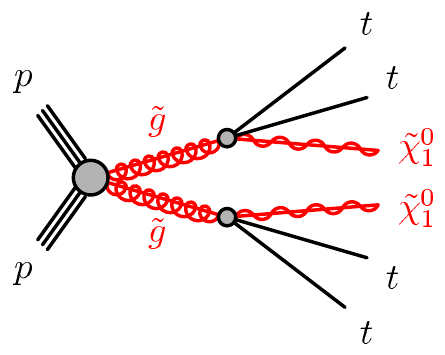}}&
\subfloat[RPV]{\includegraphics[width=5cm]{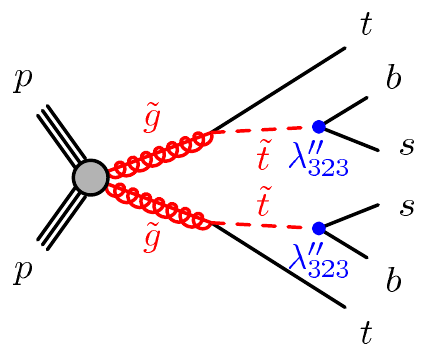}}\\
\end{tabular}
\caption{Pseudo-Feynman diagrams for the different signal models used in this search.}
\label{samples:fig:feynman_diagrams}
\end{figure}

Permitting non-zero $R$-parity-violating (RPV) couplings allows consideration of another variety of 
gluino-mediated top squark production, wherein the last step of the decay proceeds via a
baryon-number-violating interaction: $\stopone \rightarrow \bar{s} + \bar{b}$ (charge conjugates
implied). Figure~\ref{samples:fig:feynman_diagrams}(d) presents the RPV model.
Such $R$-parity-violating models give rise to final states with low missing transverse momentum.
Uniquely among the searches for strongly-produced supersymmetric particles, the current analysis selects final states with sufficiently low missing transverse momentum to be sensitive to these RPV scenarios.

The signal samples were generated using \amcatnlo~v2.2.2 interfaced to \pythia~8.186 with the A14 tune for the modelling of the parton shower (PS), hadronisation and underlying event.
The matrix element (ME) calculation was performed at tree level and includes the emission of up to two additional partons.
The PDF set used for the generation was NNPDF23LO.
The ME-to-PS matching was done using the CKKW-L prescription~\cite{Lonnblad:2001iq}, with a matching scale set to $m_{\gluino}/4$.

The gluino-mediated top squark production signal samples were generated with full simulation of the ATLAS detector, whereas  
the other signal MC samples employed the fast detector simulation.

Signal cross sections are calculated to NLO in the strong coupling constant, adding the resummation of soft gluon emission at next-to-leading-logarithm accuracy (NLO+NLL)~\cite{Beenakker:1996ch,Kulesza:2008jb,Kulesza:2009kq,Beenakker:2009ha,Beenakker:2011fu}.
The nominal cross section and the uncertainty are taken from an envelope of cross-section predictions using different PDF sets and factorisation and renormalisation scales, as described in Ref.~\cite{Borschensky:2014cia}.

\section{Event reconstruction}
\label{sec:objdef}

\subsection{Primary vertex}

Primary vertices are reconstructed using at least two charged-particle tracks with $\pT>400~\MeV$ measured by the ID~\cite{ATL-PHYS-PUB-2015-026}.
The primary vertex with the largest sum of squared track transverse momenta ($\sum{\pT^2}$) is designated as the hard scatter vertex.

\subsection{Jets}

Jets are reconstructed from three-dimensional topological clusters of calorimeter cells (topoclusters) that are noise-suppressed and calibrated to the electromagnetic scale, i.e.\ corrected for the calorimeter response to electrons and photons~\cite{PERF-2014-07}.
\emph{Small-radius jets} are built by applying the anti-$k_t$ clustering algorithm~\cite{Cacciari:2008gp}, as implemented in {\texttt{FastJet}}~\cite{Cacciari:2011ma}, with jet radius parameter $R=0.4$, to the topoclusters.
Four-momentum corrections are applied to the jets, starting with a subtraction procedure that removes the average estimated energy contributed by pile-up interactions based on the jet area~\cite{Cacciari:2007fd}.
This is followed by jet energy scale calibrations that restore the jet energy to the mean response versus particle-level simulation, using a \emph{global sequential calibration} to correct finer variations due to flavour and detector geometry and finally \emph{in situ} corrections that match the data to the MC scale~\cite{Aaboud:2017jcu}.
Only jets with $\pT>20~\GeV$ and $|\eta|<2.8$ are considered, with the exception of the \met calculation, for which jets in the range $2.8\le|\eta|<4.5$ are also used (see \Sect{\ref{sec:objdef:met}}).

To eliminate jets containing a large energy contribution from pile-up, jets are tested for compatibility with the hard scatter vertex with the \emph{jet vertex tagger} (JVT) discriminant, utilising information from the ID tracks associated with the jet~\cite{PERF-2014-03}.
Any jets with $20<\pT<60~\GeV$ and $|\eta|<2.4$ for which $\mathrm{JVT}<0.59$ are considered to originate from pile-up and are therefore rejected from the analysis.
Scale factors derived from data are applied for the simulated samples to correct the efficiency of the JVT selection.

A multivariate discriminant (MV2c10) is used to tag jets containing $B$-hadrons~\cite{ATL-PHYS-PUB-2016-012}.
This exploits the long lifetime, high decay multiplicity, hard fragmentation and large mass of $B$-hadrons.
The selected working point for the $b$-tagging algorithm~\cite{PERF-2012-04} tags $b$-jets in the pseudorapidity range $|\eta|<2.5$ with an efficiency of approximately 70\% in simulated \ttbar events, and rejects $c$-jets, $\tau$-jets and light-quark or gluon jets by factors of of 9.6, 31 and 254, respectively.
For the purposes of overlap removal, a \emph{loose $b$-tag} designation is defined using a working point with 80\% $b$-tagging efficiency.
Where $b$-tagging selections are applied, efficiency corrections measured in data are applied to simulated events, to improve modelling of the $b$-tagging efficiencies.
 
In a second jet formation step~\cite{Nachman:2014kla}, small-radius jets with $\pT>20~\GeV$ and $|\eta|<2.0$ are \emph{reclustered} using the anti-$k_t$ algorithm with radius parameters $R=1.0$ to form \emph{large-radius jets}.
The input jets are required to pass an overlap removal procedure accounting for ambiguities between jets and leptons, as discussed below.
In the leptonic control region (CR) defined in \Sect{\ref{sec:evtsel:crdefgen}}, electrons and muons may also be included in the inputs to the jet reclustering provided they satisfy $\pT>20~\GeV$ and $|\eta|<2.0$ as for standard jets.
Large-radius jets are retained for analysis if they have $\pT>100~\GeV$ and $|\eta|<1.5$.

\subsection{Electrons and photons}

Electron and photon candidates are reconstructed from clusters of calorimeter cells defined with fixed rectangular $\eta$--$\phi$ sizes and then distinguished by matching to ID tracks~\cite{PERF-2016-01,PERF-2013-04}.
A multivariate calibration is applied to correct the electron/photon energy scale~\cite{PERF-2013-05}.

Electron candidates are preselected if they have $\pT>10~\GeV$, $|\eta|<2.47$ and pass a ``Loose'' likelihood-based quality selection accounting for lateral shower shapes, hadronic shower leakage,\footnote{Energy measured in the hadronic calorimeter, within the cluster window.} hits on tracks, track--cluster matching and the number of high-threshold hits in the TRT.
Signal electrons with $\pT>20~\GeV$ are defined by requiring a ``Tight'' likelihood selection including impact parameter restrictions and the ``GradientLoose'' isolation requirement from Ref.~\cite{PERF-2016-01} in addition to the preselection.
To achieve additional rejection of background electrons from non-prompt sources, signal electron tracks must be matched to the hard scatter vertex with longitudinal impact parameter $|z_0|<0.5$~mm and transverse impact parameter significance $|d_0|/\sigma(d_0)<5$.
Corrections to the electron reconstruction and identification efficiency in simulated samples are applied using scale factors measured in data~\cite{PERF-2016-01}.

Photon candidates likewise are identified using tight criteria defined by lateral shower shapes in the first and second layers of the electromagnetic calorimeter, as well as the degree of hadronic shower leakage.
Acceptance requirements of $\pT>40~\GeV$ and $|\eta|<2.37$ are applied.

\subsection{Muons}

Muon candidates are reconstructed from tracks formed in the ID and MS, which are combined for improved precision and background rejection~\cite{PERF-2015-10}.
Stand-alone muon tracks are used to extend the muon reconstruction coverage beyond the ID acceptance in pseudorapidity (from $|\eta|=2.5$ to $|\eta|=2.7$).

Preselected muons are defined by the ``medium'' selection of Ref.~\cite{PERF-2015-10} using the number of hits on track and track quality and compatibility between the ID and MS measurements.
These must have $\pT>10~\GeV$ and $|\eta|<2.7$.
Signal muons must have a higher transverse momentum, $\pT>20~\GeV$, and satisfy the ``GradientLoose'' isolation requirement~\cite{PERF-2015-10}, as well as impact parameter matching requirements similar to those for electrons: $|z_0|<0.5$~mm and $|d_0|/\sigma(d_0)<3$.
Muon reconstruction and identification efficiencies are corrected with scale factors in simulated samples~\cite{ATL-PHYS-PUB-2015-037}.

\subsection{Overlap removal}

To avoid double counting, a procedure of overlap removal was applied to jets, photons and leptons as follows.
The electrons and muons are those passing the preselection.
\begin{enumerate}
\item If an electron and a muon share an ID track, the electron is removed and the muon is retained.
\item Photons with $\Delta R_y < 0.4$ relative to an electron or a muon are deselected.
\item Any jet that fails the loose $b$-tag selection is removed if either:
	\begin{itemize}
	\item it falls $\Delta R_y < 0.2$ from an electron; or
	\item it has no more than three tracks with $\pT>500~\MeV$, or contains an ID track matched to a muon such that $\ptjet<2\pt^\mu$ and the muon track has more than 70\% of the sum of the transverse momenta of all tracks in the jet, such that the jet resembles radiation from the muon.
	\end{itemize}
\item Any electrons or muons with $\Delta R_y < 0.4$ from a surviving jet are eliminated.
\item Finally, jets that have $\Delta R_y < 0.4$ from photons are removed.
\end{enumerate}

\subsection{Missing transverse momentum}
\label{sec:objdef:met}

The missing transverse momentum, \met, is defined as the magnitude of the negative vector sum of the transverse momenta of preselected electrons and muons, photons and jets, to which is added a ``soft term'' that recovers the contributions from other low-\pT{} particles~\cite{PERF-2014-04}.
The soft term is constructed from all tracks that are not associated with any of the preceding objects, and that are associated with the primary vertex.
In this way, the missing transverse momentum is adjusted for the best calibration of the leptons, photons and jets, while maintaining pile-up independence in the soft term.

\section{Event selection}
\label{sec:evtsel}

Target signal events for this analysis are characterised by a large jet multiplicity, beyond what is generated by high-cross-section SM processes, combined with a \met that is significantly larger than that expected purely from detector resolution.
Several signal regions are defined that select a minimum jet multiplicity and further require a large value of the ratio \metsig, where \Ht\ is the scalar sum of jet transverse momenta
\begin{equation*}
	\Ht = \sum_{j} p_{{\mathrm{T}}\\,j}^{\mathrm{jet}},
\end{equation*}
the sum being restricted to jets with $\pt>40~\GeV$, $|\eta|<2.8$.
This ratio is approximately proportional to the significance of the \met, under the assumption that the expected \met is zero and the resolution of the \met originates entirely from the stochastic variation in the jet momentum measurement.
For jets with $\pt \lesssim 1~\TeV$, the relative jet resolution scales approximately as $1/\sqrt{\pT}$.

Several auxiliary measurements are carried out in control and validation regions (VR) in order to define and constrain the major backgrounds to the analysis.
Events selected at a lower jet multiplicity are used to extract the shape of the \metsig\ distribution, which is then extrapolated to the signal regions to quantify the multijet background, as described fully in \Sect{\ref{sec:bkgest:qcdtemplate}}.
The normalisation of the \ttbar\ and \Wjets\ background components is adjusted to match data in control regions, using the procedure defined in \Sect{\ref{sec:bkgest:leptonic}}.

\subsection{Signal region definitions}

\label{sec:evtsel:srdef}

The common selection of events for all the signal regions is as follows.
To limit the contribution of SM background processes in which neutrinos are produced, leading to significant \met, events containing any preselected electron or muon following the overlap removal procedure are rejected.

Biases in the \met\ due to pile-up jets surviving the JVT selection are removed by excluding events for which a jet with $60<\pT<70~\GeV$ and $\mathrm{JVT}<0.59$ lies opposite to the \met\ ($\dphi>2.2$).
Likewise, events are rejected if they contain a jet with $\pT>50\,\GeV$ and $|\eta|<2.0$ pointing towards regions in which tile calorimeter modules were disabled. These requirements are also applied to the control regions and validation regions described later in Section~\ref{sec:evtsel:trdef}.

Subsequently, restrictions on the jet multiplicity \njet\ are imposed, depending on the analysis
channel; only jets with $\pT>50 (80)~\GeV$ and $|\eta|<2.0$ are considered as \emph{signal jets} and
therefore used in the \njet selection.
These selections are abbreviated as $j50$ ($j80$), for which the corresponding jet multiplicities are denoted \njfifty and \njeighty.
The lower and higher jet-\pt thresholds were optimised to permit sensitivity to a variety of potential SUSY mass spectra.

A threshold of $\metsig>5~\rootgev$ is the last element of the common selection.
This criterion eliminates the vast majority of SM multijet and other background events with low \met, while retaining sensitivity to a broad range of potential signals.

Next, the SRs in the two channels of the analysis are defined by a further categorisation of events.

\subsubsection{Heavy-flavour channel}

The following \njet values are considered in this channel: minimum $\njfifty \in \{8,9,10,11\}$, and minimum $\njeighty \in \{7,8,9\}$.
Motivated by the desire to achieve good sensitivity to models with differing probabilities of heavy flavour jets being produced during cascade decays, three signal regions that respectively require $\nbjet \ge 0,1,2$ are defined for each value of \njet, where the $b$-jets must have $\pt>50\,\GeV$ and $|\eta|<2.0$.

\subsubsection{Jet mass channel}

Should sparticles be produced and decay through a long decay chain, or provide enough kinetic energy to significantly boost heavy particles such as top quarks and bosons, signal events might be characterised not only by an unusually large jet multiplicity but also by the formation of large-radius jets with high masses.
The kinematic structure of SM events, by contrast, does not produce a high rate of events containing large-radius jets with mass greater than the top quark mass.

For background discrimination in this channel, a selection variable, \mjsigma, is defined to be the sum of the masses $m_j^{R=1.0}$ of the large-radius jets
\begin{equation*}
\mjsigma = \sum_j m_j^{R=1.0}
\end{equation*}
where the sum is over the large-radius jets that satisfy $\pt^{R=1.0} > 100$~\GeV\ and $|\eta^{R=1.0}| < 1.5$, as described in \Sect{\ref{sec:objdef}}.
Two thresholds for \mjsigma at $340\,\GeV$ and $500\,\GeV$, chosen following optimisation studies, define signal regions for $\njfifty \in \{8,9,10\}$, while no $j80$ SRs are defined.
As these thresholds are approximately twice and thrice the top quark mass, 
the residual irreducible backgrounds are respectively from top quark pair production in association with vector bosons and four-top processes, both of which have a very small rate.

A summary of all signal region selections is given in \Tab{\ref{tab:SRdefs}}.

\begin{table}
\renewcommand{\arraystretch}{1.1}
\centering
\begin{tabular}{l|c|c|c}
\toprule
Criterion				& \multicolumn{2}{c|}{Heavy-flavour channel}	& Jet mass channel 			\\
\midrule
Jet $|\eta|$			& \multicolumn{3}{c}{$<2.0$}												\\
\midrule
Jet \pT     			& $>50\,\GeV$			& $>80\,\GeV$				& $>50\,\GeV$					\\
\njet		         	& $\ge 8,9,10,11$ 	& $\ge 7,8,9$			& $\ge 8,9,10$				\\
\midrule
Lepton veto				& \multicolumn{3}{c}{No preselected $e$ or $\mu$ after overlap removal}	\\
\midrule
$b$-jet selection		& \multicolumn{3}{c}{$\pT>50\,\GeV$ and $|\eta|<2.0$}						\\
Large-R-jet selection	& \multicolumn{3}{c}{$\pT>100~\GeV$ and $|\eta|<1.5$}						\\
\midrule
\midrule
\nbjet					& \multicolumn{2}{c|}{$\ge 0,1,2$}			& $\ge 0$					\\
\mjsigma						& \multicolumn{2}{c|}{$\ge 0$}				& $\ge 340, 500$\,\GeV		\\
\midrule
\metsig					& \multicolumn{3}{c}{$>5\rootgev$}										\\
\bottomrule
\end{tabular}
\caption{Summary of the selection criteria for all signal regions used in this analysis.
In each column, the three selection criteria on the number \nbjet of $b$-tagged jets or the two on the the sum \mjsigma of masses of large-radius jets are applied to define separate signal regions
for each of the jet multiplicities considered. }
\label{tab:SRdefs}
\end{table}

\subsection{Control region definitions}

\label{sec:evtsel:crdefgen}
For each signal region, three control regions are used to constrain the background predictions using data, and are split into two sets.
The first set, referred to as the \emph{multijet template region} (TR) selection, maintains the same lepton veto as used in the SR, but modifies the signal jet multiplicity or \metsig selection.
Secondly, a pair of \emph{leptonic control regions} are defined, classified according to the absence or presence of a $b$-tagged jet, in which the lepton veto is replaced with a requirement on the presence of exactly one signal electron or muon (henceforth referred to merely as ``lepton'', $\ell$).

\subsubsection{Multijet template region}

\label{sec:evtsel:trdef}

Fundamental to this analysis is the extraction of an estimate of the multijet background directly from data, avoiding large theoretical uncertainties in the inclusive and differential cross-sections for these processes.
The full estimation procedure is described in \Sect{\ref{sec:bkgest:qcdtemplate}}.

Broadly, four different selections are used to evaluate the background prediction and its associated systematic uncertainties.
The shape of the full \metsig distribution (\metsig template) is measured in events containing exactly six signal jets with $\pT>50\,\GeV$ for the $j50$ signal regions and exactly five signal jets with $\pT>80\,\GeV$ for the $j80$ signal regions.
For normalisation of the template prediction, events are counted in a TR defined by the same signal jet multiplicity as the signal region, but an upper bound of $1.5~\rootgev$ on the \metsig variable.
Validation regions are defined that require seven signal jets with $\pT>50\,\GeV$ for the $j50$ signal regions and six signal jets with $\pT>80\,\GeV$ for the $j80$ signal regions, and also impose a minimum $\metsig>5~\rootgev$ threshold, as in the signal regions.
Finally, an additional validation region is defined in the range $1.5<\metsig<4.5~\rootgev$, for each signal region jet multiplicity.
The same \nbjet and \mjsigma thresholds are applied in each template and validation region as in the corresponding signal region.

\subsubsection{Leptonic control regions}

\label{sec:evtsel:crdef}

Also important is the estimation of the next two largest background processes, \ttbar and \Wjets, from MC simulation, as detailed in \Sect{\ref{sec:dataMC:background}}.
To correct for potential mismodelling of the process cross-sections and kinematics by the event generators, the normalisation for the background predictions is modified based on a simultaneous fit of the auxiliary measurements, explained in \Sect{\ref{sec:bkgest:bkgfit}}.

The leptonic control regions constraining the \ttbar and \Wjets normalisation are defined with identical selection criteria as their corresponding signal regions, apart from the following differences, summarised also in \Tab{\ref{tab:controlRegions}}:
\begin{enumerate}
\item Instead of rejecting events containing a preselected lepton, events must contain exactly one signal lepton with $\pt>20~\GeV$.
\item To prevent contamination from potential signals, events must satisfy a requirement on the transverse mass $\mt<120~\GeV$, where
	\begin{equation*}
	\mt = \sqrt{2 \pt^\ell \met \left[ 1 - \cos \left( \Delta\phi( \ptvec^\ell, \vetmiss ) \right)
  \right]}.
	\end{equation*}
\item To increase the number of selected events, the minimum signal jet multiplicity \njet is reduced by one from the corresponding signal region. However, if the lepton satisfies the \pT and $\eta$ requirements imposed on signal jets, then it is treated as a signal jet for the purposes of this selection. This reflects the main mechanism by which \ttbar and \Wjets events pass the signal region selection: misidentification of an electron or hadronically-decaying tau lepton as a jet, which can increase the jet multiplicity. Events with leptons which are unreconstructed as they lie outside of detector acceptance can also contribute to the signal regions, but are a subdominant contribution.
\item Events consistent with \Wjets and \ttbar production are separated by means of the \nbjet selection; the \Wjets CR requires $\nbjet = 0$ while the \ttbar CR requires $\nbjet \geq 1$.
\item The \metsig threshold is lowered from $5~\rootgev$ to $3~\rootgev$ or $4~\rootgev$ when it is necessary to increase the statistical precision of the measurement. The \metsig thresholds are specified in \Tab{\ref{tab:CR_metsigCuts}}.
\end{enumerate}

\begin{table}[htb]
\centering
\begin{tabular}{@{}ll@{}}
\toprule
\multicolumn{2}{l}{Control regions} \\ \midrule
Lepton multiplicity					&  Exactly one signal $e$ or $\mu$\\
Lepton \pt							&  $>20$~\GeV\ \\
\mt									&  $<120$~\GeV\\
Jet \pt, $|\eta|$					&  Same as SR\\
Number of jets including lepton		&  $\njet^\mathrm{SR}-1$\\
$b$-jet multiplicity					&  $=0$ (\Wjets) or $\geq 1$ (\ttbar)\\
\mjsigma							&  Same as SR \\
\metsig								&  $> 3, 4, 5$\,\rootgev \\
\bottomrule
\end{tabular}
\caption{Definition of the leptonic control regions, used to normalise the
\ttbar and \wjets backgrounds. In the control regions, the lepton is recast as
a jet if it satisfies the same kinematic criteria as the jets. Such
leptons contribute to the \metsig (through \Ht) and also \mjsigma. }
\label{tab:controlRegions}
\end{table}

\begin{table}[htb]
\renewcommand{\arraystretch}{1.1}
\centering
\begin{tabular}{l|c|c|c}
\toprule
Signal channel
	& \multicolumn{2}{c|}{Minimum SR \njet}	&	\metsig threshold \\
\midrule
\multirow{4}{*}{Heavy-flavour channel} 
  & Jet $\pT>50$\,\GeV & Jet $\pT>80$\,\GeV &                \\
  & 8, 9               & 7                  & $> 5$ \rootgev \\
  & 10                 & 8                  & $> 4$ \rootgev \\
  & 11                 & 9                  & $> 3$ \rootgev \\
\midrule
\multirow{4}{*}{Jet mass channel}
  & $\mjsigma > 340$\,\GeV & $\mjsigma > 500$\,\GeV &                \\
  & 8                 & -                 & $> 5$ \rootgev \\
  & 9                 & 8                 & $> 4$ \rootgev \\
  & 10                & 9, 10             & $> 3$ \rootgev \\
\bottomrule
\end{tabular}
\caption{
The \metsig thresholds for the control regions corresponding to each signal region.
In each case, the same \metsig threshold is used for both the \Wjets and \ttbar control regions.
}
\label{tab:CR_metsigCuts}
\end{table}

\section{Background estimation techniques}
\label{sec:bkgest}

\subsection{Multijet template estimation}
\label{sec:bkgest:qcdtemplate}

Accurate modelling of multijet processes by performing QCD calculations involving high multiplicity multi-leg matrix elements is difficult.
This is compounded by the challenges of reproducing events populating the tails of the detector response, representative of the high-\met events selected in this analysis.
Hence, to confidently estimate the multijet background component, which makes up 50--70\% of the total SM expectation, the prediction is based on direct measurements in data.

The strategy used in this analysis is based on the observation that the \metsig spectrum for selected multijet events is primarily determined by the calorimeter response to jets, which is approximately independent of how the total jet transverse energy \Ht is partitioned between the jets.
Thus, the \metsig spectrum measured in events with a lower jet multiplicity does not greatly differ from that observed in events with a high jet multiplicity.
A \emph{template} for the multijet \metsig distribution can thus be extracted in a selection complementary to the signal region, specifically the template regions defined in \Sect{\ref{sec:evtsel:trdef}}.
At larger values of \metsig, it is necessary to subtract from the data the expected contributions due to SM processes producing neutrinos. For the \metsig threshold used in the SRs, such contributions comprise approximately 10\% to 50\% of the total. These predictions are determined from MC simulation.
This template also accounts for smaller background contributions from \ttbar production with fully hadronic decays as well as $\gamma+\mathrm{jets}$.

By the logic above, the multijet prediction $n_\mathrm{multijet}$ for the number of events with $b<\metsig<c$ for a SR based on a TR can be written as follows:
\begin{align*}
	n_\mathrm{multijet}^{\mathrm{SR}, b<\metsig<c} &= 
		\frac{n_\mathrm{multijet}^{\mathrm{SR}, \metsig<a}}{n_\mathrm{multijet}^{\mathrm{TR}, \metsig<a}}
			\cdot n_\mathrm{multijet}^{\mathrm{TR}, b<\metsig<c} \\
	&= 	\frac{n_\mathrm{multijet}^{\mathrm{SR}, \metsig<a}}{n_\mathrm{multijet}^{\mathrm{TR}, \metsig<a}}
			\cdot \left(n_\mathrm{obs}^{\mathrm{TR}, b<\metsig<c} - n_{\mathrm{MC} \nu}^{\mathrm{TR}, b<\metsig<c} \right).
\end{align*}
The normalisation of the template is fixed in the range $\metsig<a$ such that $a<b<c$, which is entirely dominated by multijet events.
In the template region, the observation in data is denoted $n_\mathrm{obs}$, while the predicted number of events with neutrinos is written $n_{\mathrm{MC}\nu}$.

While the exact division of \Ht among the multiple jets in a single event does not significantly influence the template independence, the distribution of \Ht itself is forced higher as the \njet requirements are made more stringent.
This implies the existence of an indirect correlation between the \metsig and the jet multiplicity, which challenges the earlier assumption of template independence.
It is therefore necessary to extract the multijet template in several bins of \Ht in order to remove the subdominant residual dependence of \metsig on \Ht.
The lower bin boundaries are set at 0, 600, 900 and 1200~\GeV, which was found to be sufficient to remove the dependence of the template on \Ht.
Predictions for each bin are derived independently and summed to obtain the total SR expectation.

The dependence of the template prediction on pile-up was studied in detail.
While the width of the \metsig distribution itself shows a correlation with the amount of pile-up, both due to the growth of the jet resolution and the influence of additional jets present in the event, increases in the amount of pile-up do not worsen the closure of the template prediction, \mbox{i.e.} the ability of the method to correctly predict the multijet background in validation regions free of signal.
This demonstrates that the template method accurately captures the variation in the \metsig spectrum under changing LHC conditions, and that there is no observable bias in the signal region acceptance for a given level of pile-up.

Other potential influences on the closure of the template such as the heavy-flavour composition were also studied carefully.
The most important of these effects are kinematic variations between the template and signal regions, and differences in the number of heavy-flavour jets in the two event selections.
While no major changes in the prediction were observed in these checks, systematic uncertainties that estimate the sensitivity of the template prediction to these variables are assessed.

Kinematic differences are covered by comparing the nominal estimate to the prediction extracted using an alternative \Ht-binning strategy, setting the bin boundaries instead at 0, 600, 800, 1000, 1200 and 1400~\GeV.
The resulting uncertainty is typically 5--10\% in the SRs.

An uncertainty due to jet flavour composition is determined as the difference between the nominal estimate, which assumes an identical flavour composition between the TR and SRs, and a $\chi^2$ fit that interpolates between the nominal estimate and a \emph{flavour-split} template estimate.
The flavour-split template prediction is produced by separating the template and signal regions into two bins, one requiring exactly as many $b$-jets as in the SR \nbjet selection, and the other requiring at least one more.
A $\chi^2$ fit to data in the validation regions is then used to linearly combine the nominal and flavour-split templates.
The resulting combined template is used as a basis for comparison to the nominal prediction.
This procedure ensures that an appropriate uncertainty is estimated if the nominal estimate is significantly different from the best fit; if the na\"ive flavour-split estimate describes the data poorly, this does not result in an overestimate of the uncertainty.
For the jet mass channel, this uncertainty ranges from 3\% to 6\%.
It is larger in the heavy-flavour channel: at most 20\% in the tightest selections, and up to 12\% elsewhere.

Finally, to account for other potential sources of mismodelling, an overall closure uncertainty is computed.
This is defined as the maximal relative difference between the template prediction and the observation in data for the VRs defined in \Sect{\ref{sec:evtsel:trdef}}, either with a lower jet multiplicity or a reduced \metsig value.
The template closure is checked in a VR at a lower jet multiplicity but with the same $\metsig>5\rootgev$ threshold as in the SR, or in several bins of \metsig:
\begin{equation*} 
  \metsig \in (1.5,2.0), \, (2.0,3.0), \, (3.0,4.0)~\rootgev.
\end{equation*}

Example distributions of \metsig in the lower-jet-multiplicity VRs are shown in Figure~\ref{fig:dataMC_TRVR_Summary}. 
The degree of closure varies, generally ranging between 8\% and 12\% and extending to 30\% for regions with the fewest events.

\newcommand{\uncertaintyBand}{The blue hatched band indicates the quadrature sum of the statistical uncertainty from MC simulated samples and the various systematic uncertainties in the background prediction.}
\newcommand{\statBand}{The blue hatched band indicates the statistical uncertainty from MC simulated samples.}
\newcommand{\scaledBackgrounds}{The backgrounds are scaled by the normalisation factors extracted from the fit, described in Section~\ref{sec:bkgest:bkgfit}.}
\newcommand{\benchmarkModels}{The dashed lines labelled `pMSSM' and `2-step' refer to benchmark signal points -- a pMSSM slice model with $(m_{\gluino}, m_{\tilde{\chi}_1^\pm}) = (1400, 200)$\,\GeV~and a cascade decay model with $(m_{\gluino}, m_{\tilde{\chi}_1^0}) = (1400, 200)$\,\GeV.}

\newcommand{\APPROVAL}{}
\newcommand{\plotsuffix}{_systBand_combined\APPROVAL}
\newcommand{\plotsuffixfit}{_fitted_systBand_combined\APPROVAL}
\newcommand{\panelnoredarrow}{The lower panels show the ratio of the observed data to the total SM background.}
\newcommand{\redarrow}{Red arrows indicate data points for which the error bar does not intersect the ratio plot.}

\begin{figure}[htb]
\centering
\begin{tabular}{cc}
\subfloat[$\njfifty=7$, $\nbjet\geq0$]{\includegraphics[width=0.45\textwidth]{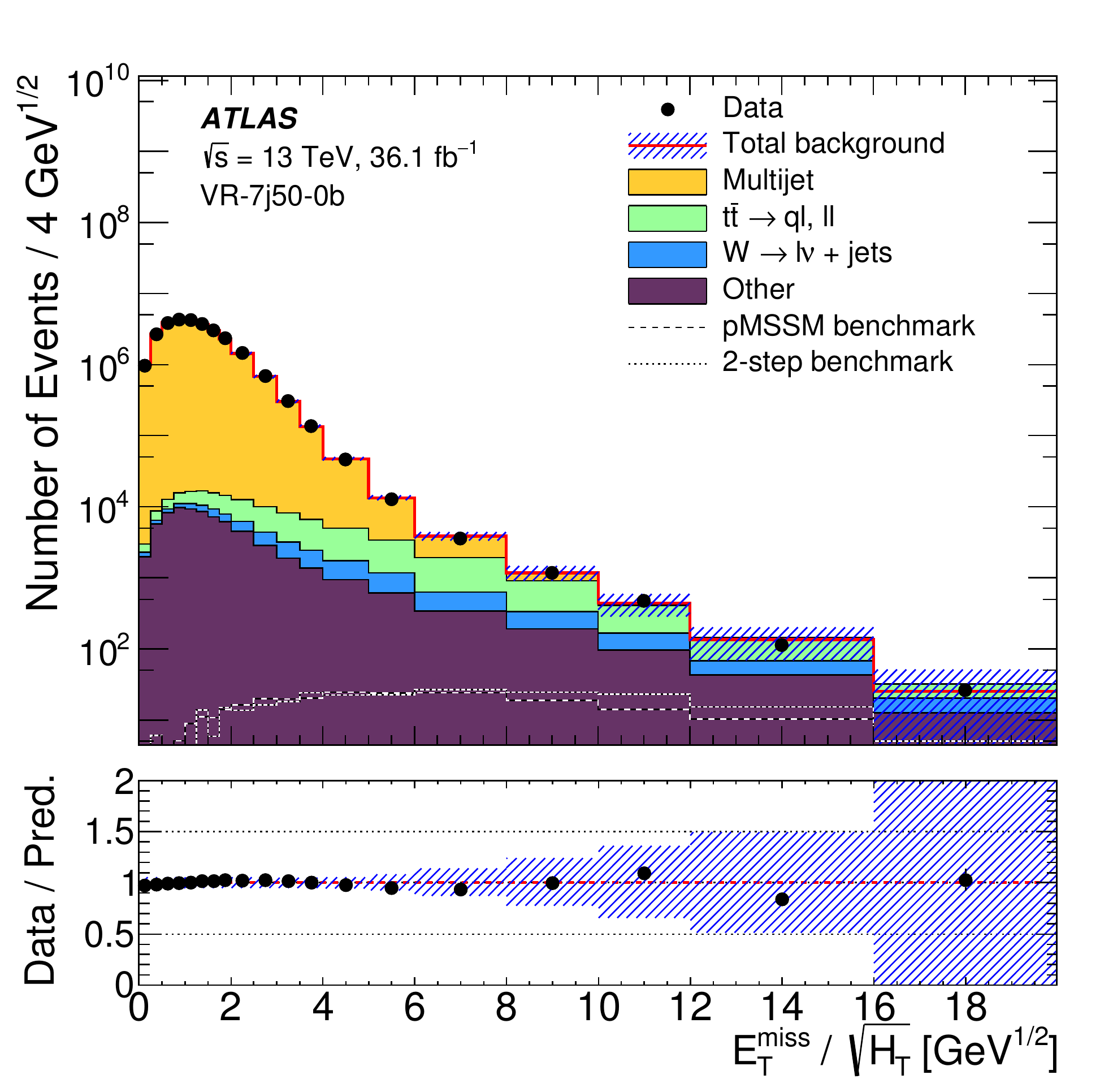}}\hspace{4ex}&
\subfloat[$\njfifty=7$, $\mjsigma > 500~\GeV$]{\includegraphics[width=0.45\textwidth]{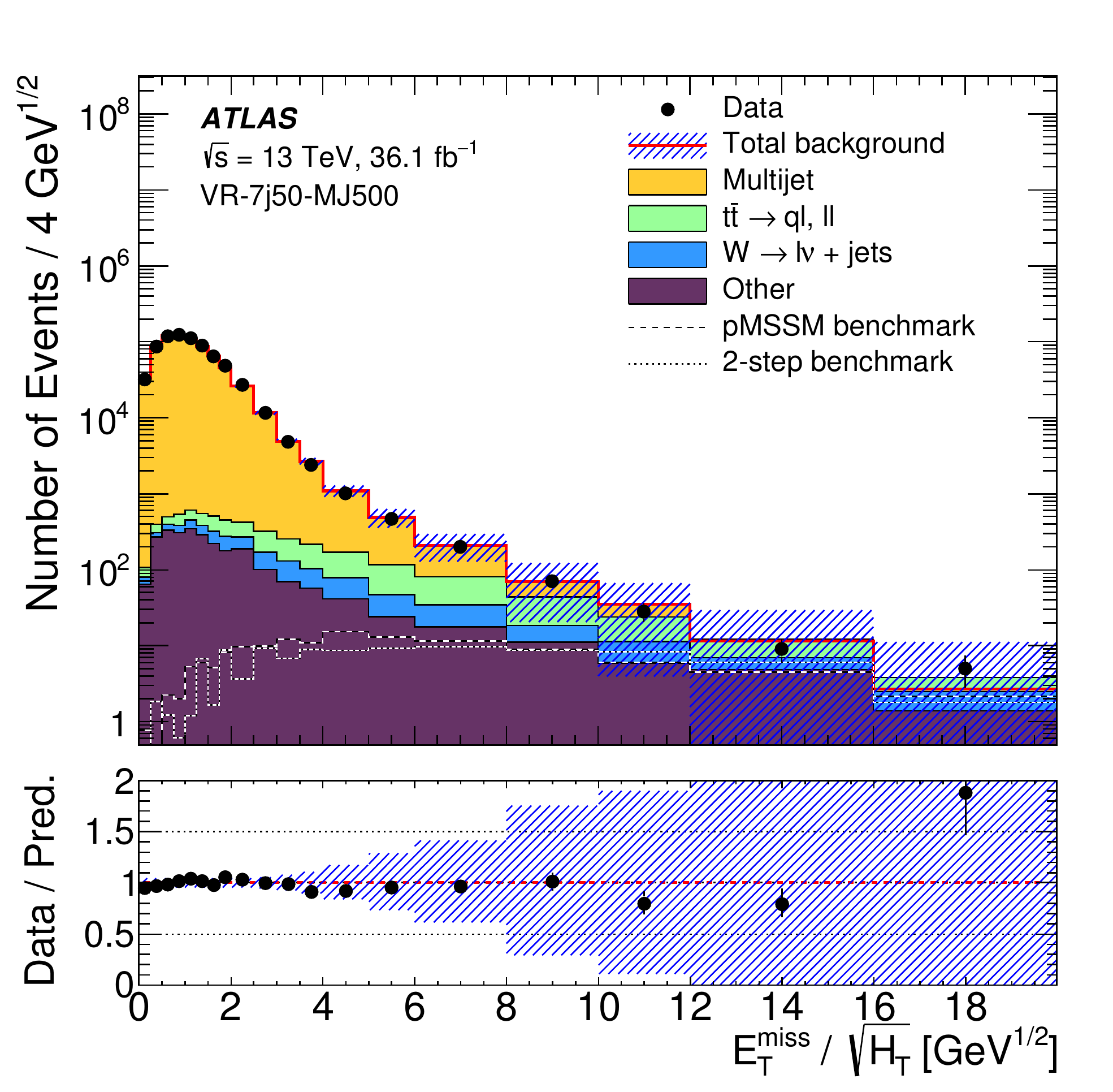}}\\
\end{tabular}
\caption{Distributions of the \metsig for events in the validation regions for the 50\,\GeV\ jet \pT threshold in the heavy-flavour channel (a) and a \mjsigma selection (b). 
\uncertaintyBand~\benchmarkModels~\panelnoredarrow}
\label{fig:dataMC_TRVR_Summary}
\end{figure}

\clearpage

\subsection{Leptonic background estimates}
\label{sec:bkgest:leptonic}

All background contributions from processes in which \Wln or \Znn decays produce neutrinos, including single or pair production of top quarks and electroweak vector bosons, are estimated using MC simulation.
The two largest of these, \ttbar and \Wjets, are responsible for 20--45\% and up to 10\% of the SM background respectively.
Other processes, such as \Zjets, single top and diboson production collectively make up no more than 12\% of the total SR expectation.
As such, corrections to the size of the \ttbar and \Wjets background components, together with the multijet template estimate previously described, provide a sufficiently accurate background prediction for this search.
For each of the \ttbar, \Wjets and multijets background processes, a normalisation factor $\mu$ is determined, based on a likelihood fit described in \Sect{\ref{sec:bkgest:bkgfit}}.

Control regions defined as in \Sect{\ref{sec:evtsel:crdef}} provide enriched samples of events from the relevant processes, in a kinematic region close to the signal selection.
The purity of the CRs is around 85\% for \ttbar and typically 25--50\% for \Wjets.
As only these two processes contribute substantially to the CR populations, this level of purity is adequate to constrain the normalisations for both well.

Distributions of the number of jets ($\pt>20~\GeV$, $|\eta|<2.8$) are shown in Figure~\ref{fig:controlRegions_nGoodJets} for a selection of the \ttbar and \Wjets CRs.

\begin{figure}
  \centering
  \begin{tabular}{cc}
    \subfloat[$\njfifty\geq7$, $\nbjet=0$]{\includegraphics[width=0.38\linewidth,clip,trim=0 0 0 3ex]{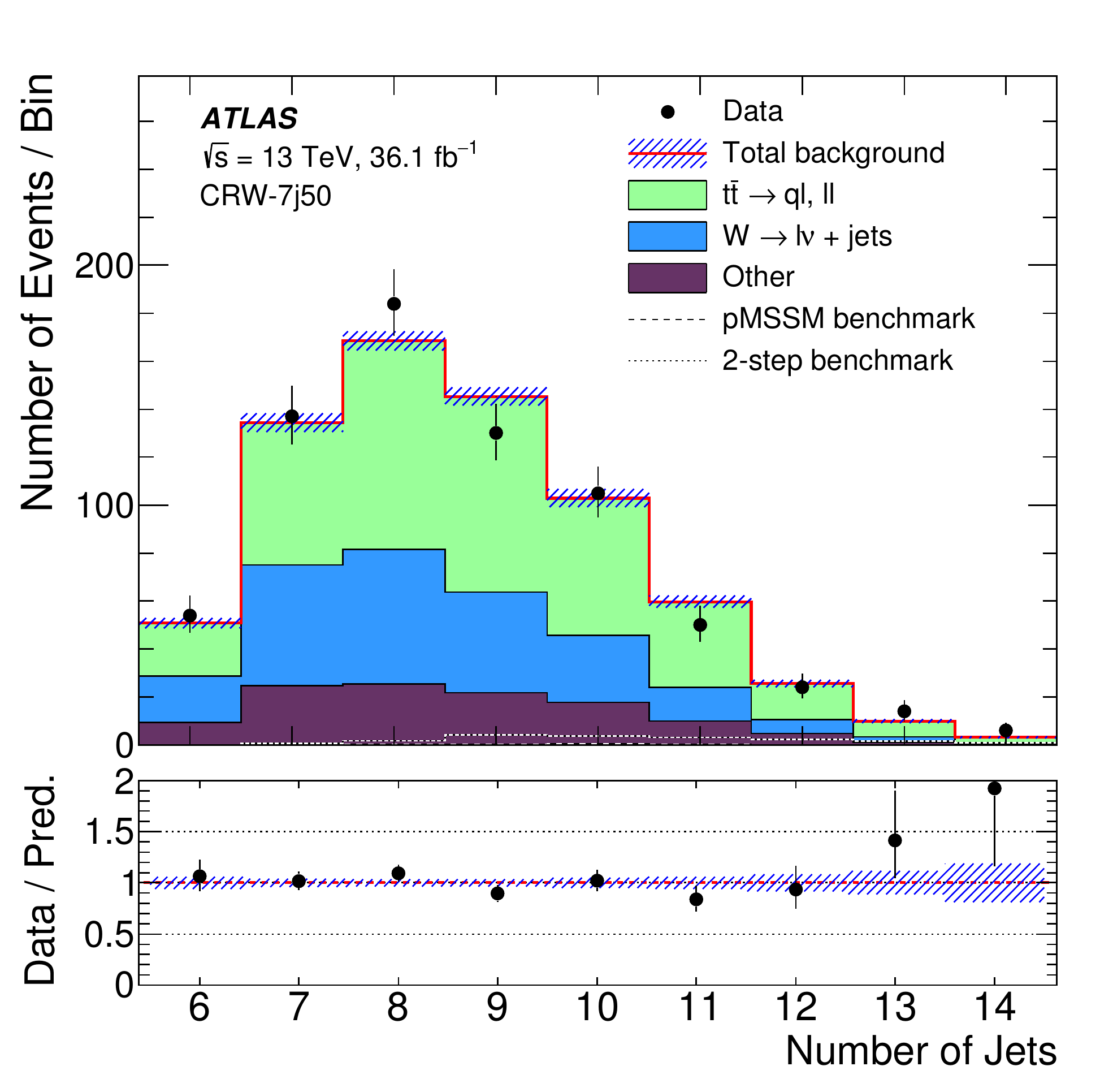}} &
    \subfloat[$\njfifty\geq7$, $\nbjet\geq1$]{\includegraphics[width=0.38\linewidth,clip,trim=0 0 0 3ex]{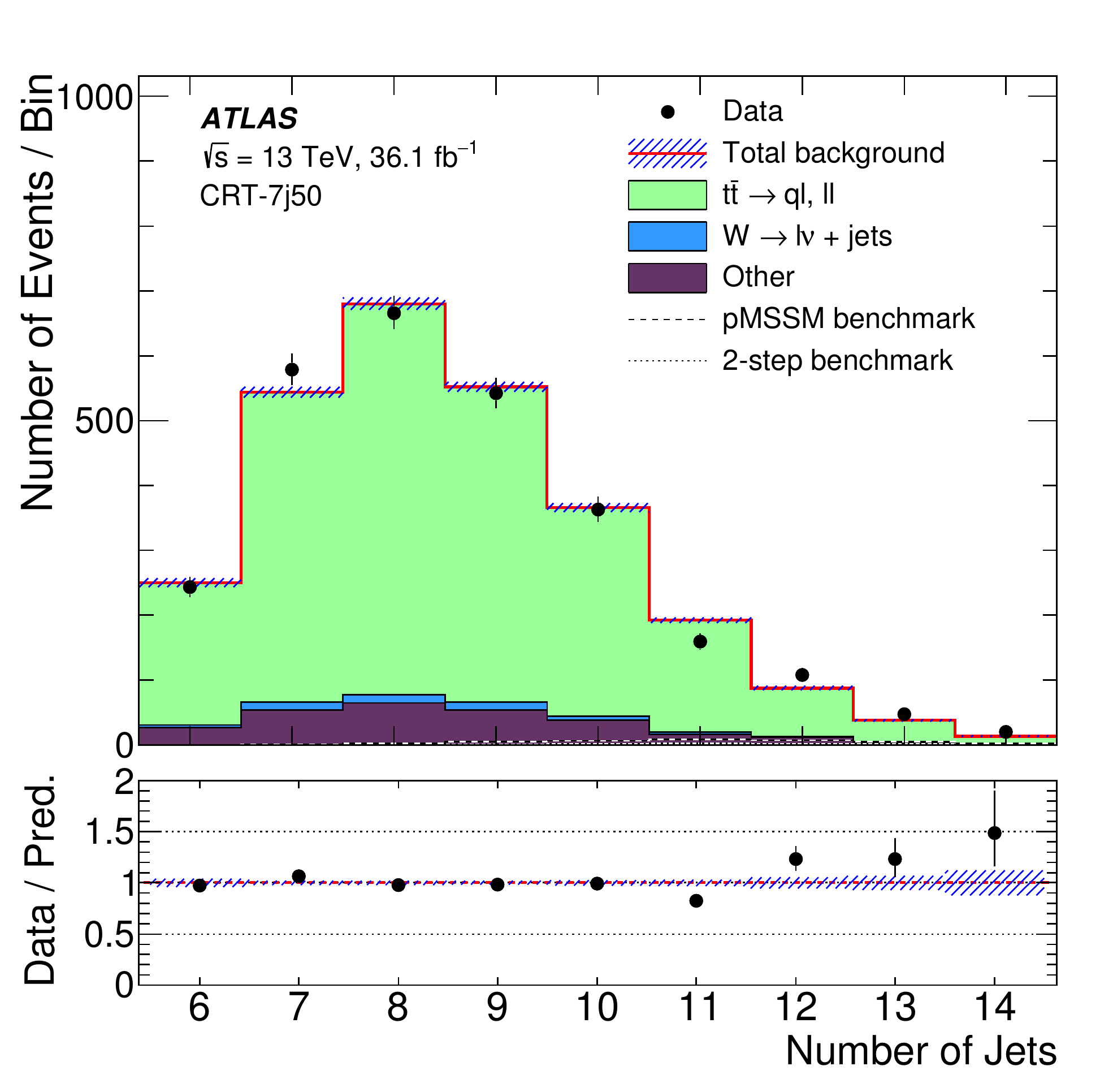}} \\    
    \subfloat[$\njeighty\geq6$, $\nbjet=0$]{\includegraphics[width=0.38\linewidth,clip,trim=0 0 0 3ex]{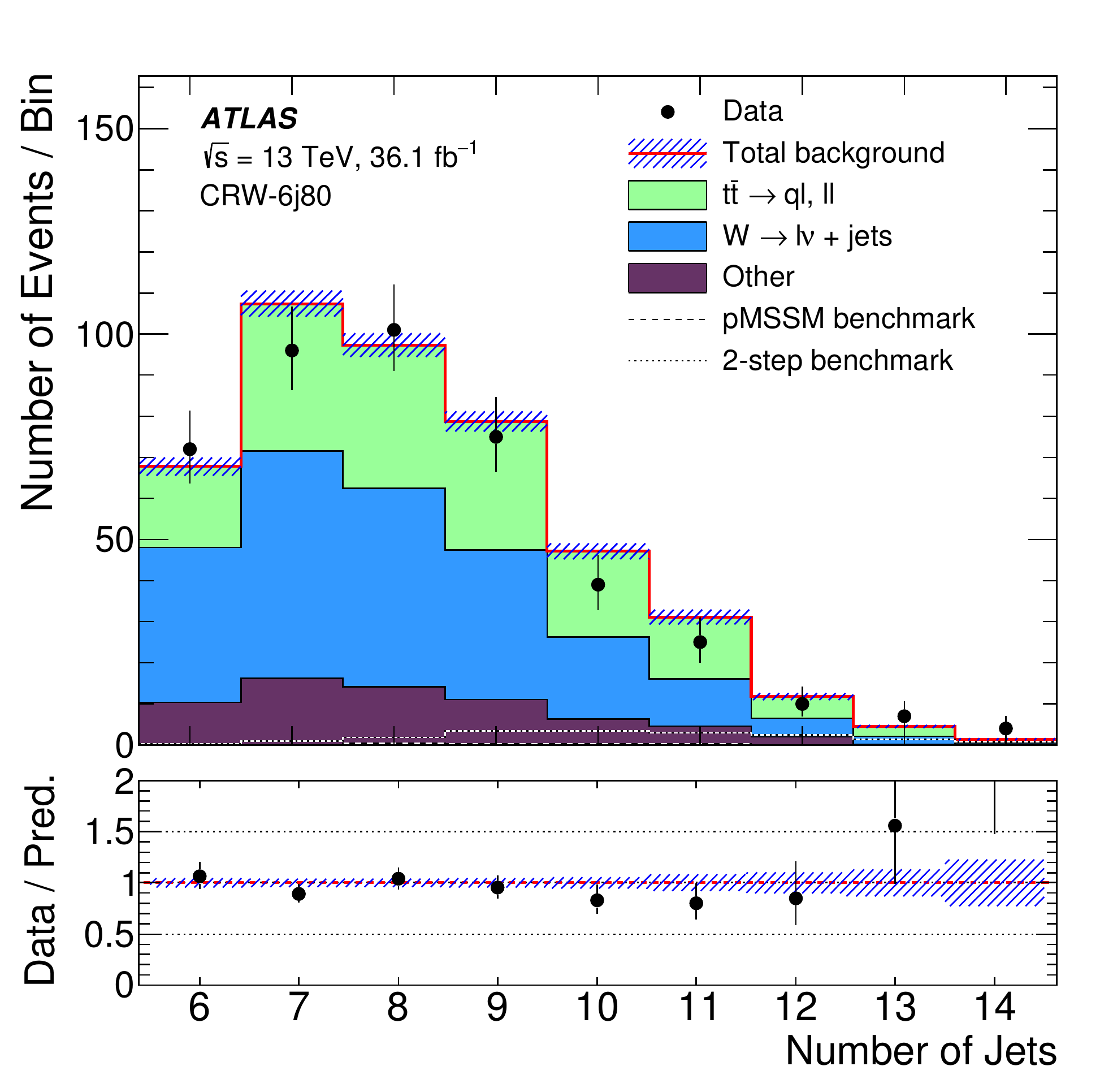}} &
    \subfloat[$\njeighty\geq6$, $\nbjet\geq1$]{\includegraphics[width=0.38\linewidth,clip,trim=0 0 0 3ex]{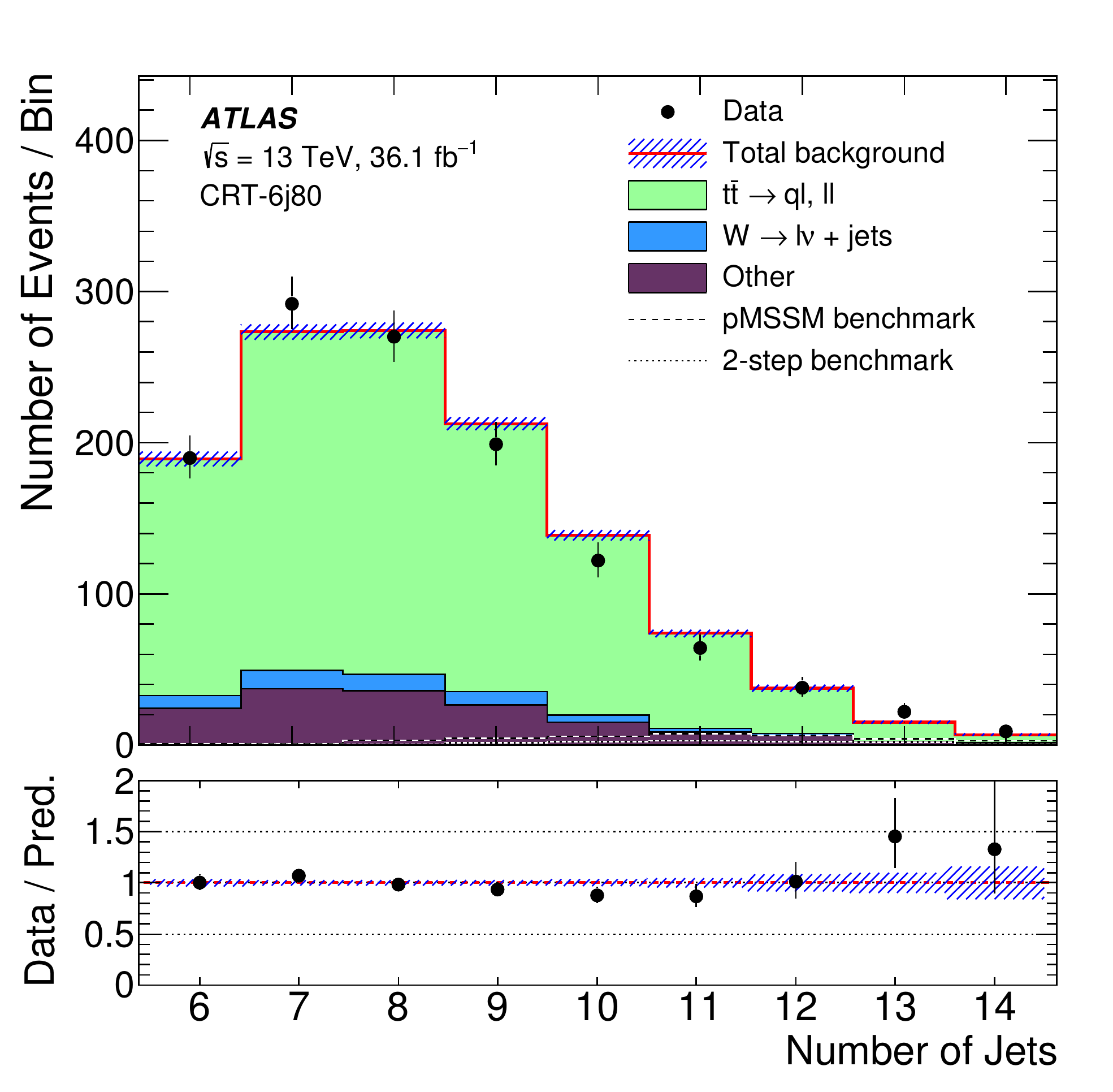}} \\    
    \subfloat[$\njfifty\geq7$, $\mjsigma>340~\GeV$, $\nbjet=0$]{\includegraphics[width=0.38\linewidth,clip,trim=0 0 0 3ex]{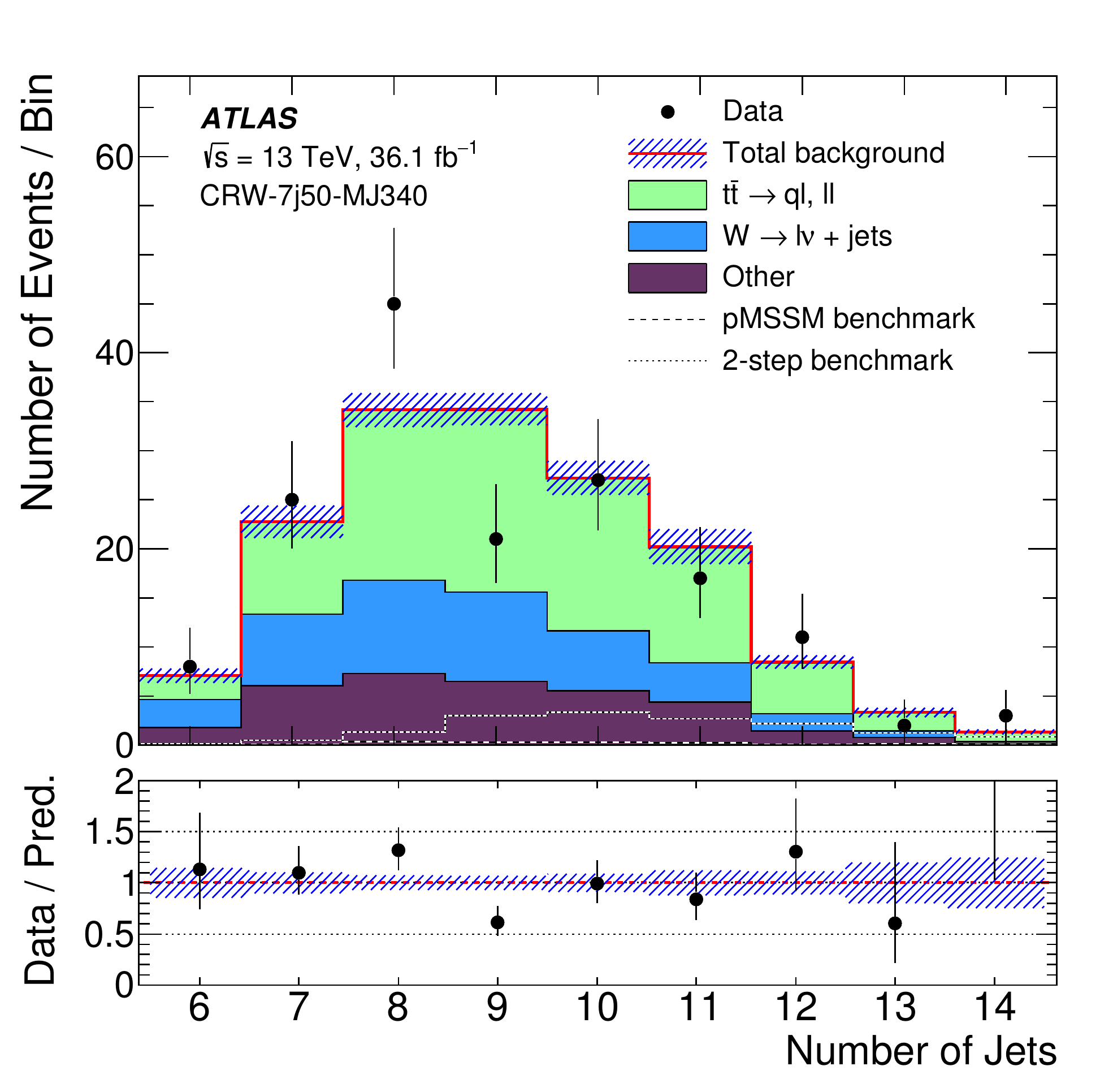}} & 
    \subfloat[$\njfifty\geq7$, $\mjsigma>340~\GeV$, $\nbjet\geq1$]{\includegraphics[width=0.38\linewidth,clip,trim=0 0 0 3ex]{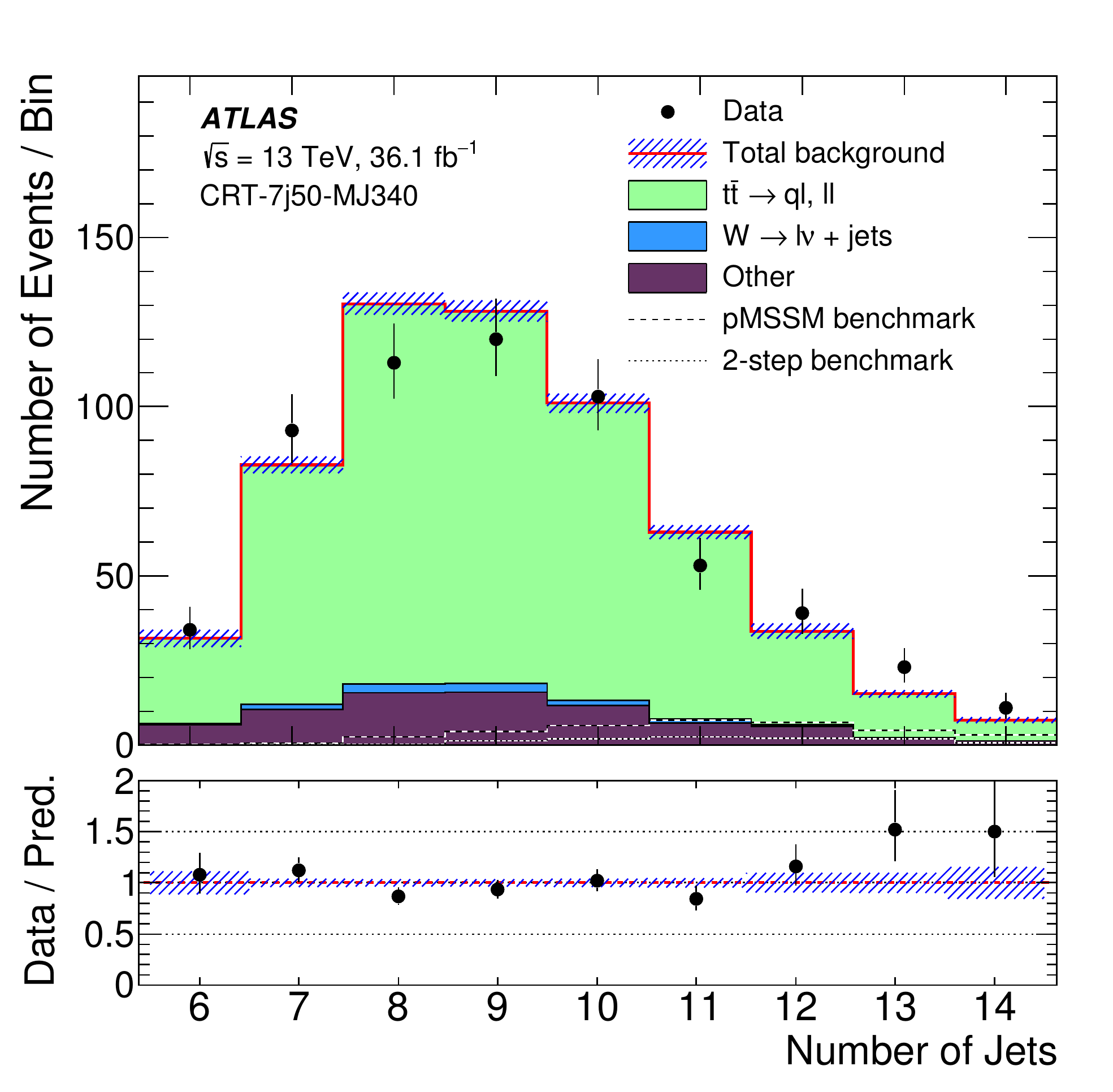}} \\    
  \end{tabular}
  \caption{The distribution of the number of jets observed in the \wjets (left) and \ttbar (right) control regions with
  the lowest jet multiplicities. 
\scaledBackgrounds~\statBand~\benchmarkModels~\panelnoredarrow
}
  \label{fig:controlRegions_nGoodJets}
\end{figure}

\subsection{Combined background fits}

\label{sec:bkgest:bkgfit}

For each background process constrained by the fit, an unconstrained normalisation factor $\mu_b,\; b \in \{\ttbar,W,\mathrm{multijet}\}$ is defined, such that $\mu_b=1$ implies consistency with the nominal MC cross-sections for \ttbar and \Wjets.
The normalisation factor $\mu_\mathrm{multijet}$ allows the MC subtraction applied in the template estimate to be corrected by the CR measurements, and to be modified coherently with any systematic variations applied to the MC simulation.

A likelihood is then constructed for the ensemble of measurements in the control regions as the product of Poisson distributions whose means are specified by the nominal MC estimate for that region, including the free normalisation factors $\mu_b$~\cite{Baak:2014wma}.
For $\mu_{\ttbar}$ and $\mu_W$, the corresponding leptonic control regions provide the constraints.
The 6-jet ($j50$) or 5-jet ($j80$) template region is treated as another control region in the fit such that $\mu_\mathrm{multijet}$ is coupled to any modifications of $\mu_{\ttbar}$ and $\mu_W$.
If $\mu_{\ttbar}=\mu_W=1$, then $\mu_\mathrm{multijet}=1$ by construction, as the same region is used to derive the nominal multijet estimate.
The systematic uncertainties (see Section~\ref{sec:stats:systematics}) are implemented in the form of Gaussian-constrained nuisance parameters modifying the Poisson mean of each background component contributing to the estimate in a given signal or control region.

Minimisation of the likelihood (profiling) fixes the values of, and uncertainties in,
$\mu_b$, which can then be combined with the MC and template predictions
to obtain the total background prediction in the signal region.
The compatibility of the background prediction and SR observation is computed in the form of a $p$-value ${\mathrm{CL}}_{\mathrm{b}}$ which is the probability of an upwards fluctuation from the SR prediction no larger than that observed in data, given the background model.

\section{Statistical procedures}
\label{sec:stats}

\subsection{Systematic uncertainties}

\label{sec:stats:systematics}

Systematic uncertainties affecting this analysis are grouped into the three following sources.

Uncertainties from experimental sources include those in identification and reconstruction efficiencies, as well as energy and momentum scales and resolutions. They are are assessed for all simulated event samples.
Efficiency uncertainties are considered for hard-scatter jet selection, flavour tagging and selection of electrons and muons.
Of these, only flavour-tagging uncertainties have a non-negligible effect on the total background expectation in the signal regions; at most 4\% in the heavy-flavour-enriched SRs.

The energy/momentum uncertainties affect jets, electrons, muons and photons, and are also propagated to the missing transverse momentum.
Jet energy scale and resolution systematic uncertainties contribute 6--12\% to the uncertainty in the SR yields.
The soft term of the \met also has its own associated uncertainties, which in the jet mass channel may have up to an 8\% effect.
In this category also fall the uncertainty in the total integrated luminosity considered for analysis as well as in the total inelastic $pp$ cross-section, which affects the simulation of pile-up (<1\%).

Theoretical uncertainties in the event generation affect both the background and signal MC samples. These are assessed by varying the matrix element and parton shower generators used, or by modifying scales (renormalisation, factorisation, resummation, matching) involved in the process calculations.
Variation in the degree of additional QCD radiation accompanying \ttbar production is the single largest source of uncertainty in the SRs (10--25\%); parton shower uncertainties play a subdominant role, typically being half as large as, but occasionally comparable to, the radiation systematic uncertainty.
Constant uncertainties of 30\% and 50\% respectively are applied to the normalisation of diboson production and top quark pair production in association with vector bosons, and have an overall negligible effect on the analysis results.

As described in \Sect{\ref{sec:bkgest:qcdtemplate}}, uncertainties in the multijet background estimates are assessed, where kinematic and flavour differences between the template and signal regions are considered.
An additional overall systematic uncertainty is ascribed for general non-closure of the template prediction.
Apart from in the jet mass channel SRs, where the kinematic and flavour uncertainties are at most 3\%, and in the most statistically limited SRs, the uncertainties from the three sources are similar in magnitude.
Where the statistical precision is poorer, fluctuations can drive the non-closure uncertainty up to 18\%.

\subsection{Hypothesis testing}

\label{sec:stats:hypotest}

For the interpretation of the signal region observations, the likelihood fits for background estimation (\Sect{\ref{sec:bkgest:bkgfit}}) are extended to perform two forms of hypothesis tests using a profile-likelihood-ratio test statistic~\cite{statforumlimits}, quantifying the significance of any observed excesses or the lack thereof.
The discovery test discriminates between the null hypothesis stating that the SR measurement is consistent with only SM contributions and an alternative hypothesis postulating a positive signal.
Conversely, any given signal model can be examined in an exclusion test of the signal-plus-background hypothesis, where an observation significantly smaller than the combination of SM and SUSY processes would lead to rejection of the signal model.

Taking into account all background predictions, normalisation factors and systematic uncertainties, the fit is implemented by including the SR in the ensemble of measurements and adding an additional signal component solely in the SR.
Using a profile-likelihood-ratio test, the discovery $p$-value $p_0$, corresponding to the probability of an upward fluctuation in the absence of any signal, can be determined.
This configuration also permits an upper limit on the visible signal cross-section to be set for an arbitrary signal, where it is assumed that the signal contamination in the control regions is negligible.

Exclusion testing of a chosen signal model proceeds similarly, but a signal component is allowed in all control regions as well as the signal region, to correct for potential signal contamination (which has been verified to be small).
Theoretical and experimental systematic uncertainties in the signal MC simulation are included in the fit.
A profile-likelihood-ratio test is then made of the compatibility between the best-fit $\mu_\mathrm{signal}$ from data and the nominal signal hypothesis, corresponding to a signal strength $\mu_\mathrm{signal}=1$.
This provides the exclusion $p$-value $p_1$.
Points in the SUSY parameter space are considered excluded if the ${\mathrm{CL}}_{\mathrm{s}}$ parameter, computed as $p_1/(1-{\mathrm{CL}}_{\mathrm{b}})$, is smaller than 0.05~\cite{clsread}.
This protects against spurious exclusion of signals due to observing SR event counts significantly smaller than those predicted.
While not strictly defining a frequentist confidence level, these are referred to as 95\% confidence level \mbox{CL} limits.

\section{Results and interpretation}
\label{sec:result}

The expected and observed event counts in the leptonic control regions are evaluated and normalisation factors derived.
In general, the \ttbar normalisation is close to one for lower jet multiplicities but may be as small as 0.71 for high jet multiplicities.
For $\mu_{W}$, the range is typically 0.3--0.6.
Correspondingly, $\mu_\mathrm{multijet}$ is corrected upwards by up to 24\%.

\newcommand{\HFPlotDir}{postFit}

Signal region yields as observed in data are summarised in \Tab{\ref{tab:fitResults}}.
These are illustrated graphically in Figure~\ref{fig:regionSummarySR_postFit}.
The most significant difference from the SM prediction is a deficit in the 9j MJ500 region with a statistical significance of $1.8\sigma$ and a corresponding $p$-value $(1-{\mathrm{CL}}_{\mathrm{b}})$ of 0.04. Similar deficits are observed in
the other MJ SRs, but the large overlap between these SRs implies that the deficits are strongly correlated.

\begin{table}[htb]
\centering
\renewcommand\arraystretch{1.2}

\begin{tabular}{@{}>{\small}llr@{${}\pm{}$}lr@{${}\pm{}$}lr@{${}\pm{}$}lr@{}}
\toprule

\multicolumn{2}{l}{\multirow{2}{*}{Signal region}}
& \multicolumn{6}{c}{Fitted background}
& \multirow{2}{*}{Obs events} \\
\cline{3-7}
& & \multicolumn{2}{c}{Multijet}
& \multicolumn{2}{c}{Leptonic}
& \multicolumn{2}{c}{Total} &\\
                                       
\midrule
\multirow{3}{*}{$N_{\mathrm{jet}}^{50} \geq 8$}
& $ N_{ { \bjet }} \geq 0$ & 622 & 42 & 570 & 140 & 1190 & 140 & 1169 \\ 
& $ N_{ { \bjet }} \geq 1$ & 460 & 50 & 430 & 110 & 890 & 140 & 856 \\ 
& $ N_{ { \bjet }} \geq 2$ & 196 & 39 & 226 & 57 & 422 & 81 & 442 \\ 
\midrule
\multirow{3}{*}{$N_{\mathrm{jet}}^{50} \geq 9$}
& $ N_{ { \bjet }} \geq 0$ & 96 & 11 & 98 & 24 & 194 & 28 & 185 \\ 
& $ N_{ { \bjet }} \geq 1$ & 84 & 15 & 76 & 20 & 160 & 31 & 135 \\ 
& $ N_{ { \bjet }} \geq 2$ & 39 & 12 & 42.5 & 9.5 & 82 & 19 & 76 \\ 
\midrule
\multirow{3}{*}{$N_{\mathrm{jet}}^{50} \geq 10$}
& $ N_{ { \bjet }} \geq 0$ & 15.1 & 3.0 & 18.3 & 3.9 & 33.5 & 5.1 & 26 \\ 
& $ N_{ { \bjet }} \geq 1$ & 15.3 & 3.7 & 14.7 & 3.3 & 30.0 & 5.9 & 23 \\ 
& $ N_{ { \bjet }} \geq 2$ & 7.6 & 3.1 & 8.4 & 1.8 & 16.0 & 4.2 & 15 \\ 
\midrule
\multirow{3}{*}{$N_{\mathrm{jet}}^{50} \geq 11$}
& $ N_{ { \bjet }} \geq 0$ & 2.54 & 0.76 & 2.4 & 1.2 & 4.9 & 1.2 & 7 \\ 
& $ N_{ { \bjet }} \geq 1$ & 2.88 & 0.84 & 2.1 & 1.4 & 5.0 & 1.3 & 6 \\ 
& $ N_{ { \bjet }} \geq 2$ & 1.49 & 0.72 & 1.6 & 1.5 & 3.1 & 1.5 & 4 \\ 
\midrule
\midrule
\multirow{3}{*}{$N_{\mathrm{jet}}^{80} \geq 7$}
& $ N_{ { \bjet }} \geq 0$ & 282 & 32 & 253 & 69 & 535 & 74 & 486 \\ 
& $ N_{ { \bjet }} \geq 1$ & 219 & 28 & 183 & 60 & 402 & 74 & 343 \\ 
& $ N_{ { \bjet }} \geq 2$ & 100 & 17 & 91 & 34 & 191 & 44 & 160 \\ 
\midrule
\multirow{3}{*}{$N_{\mathrm{jet}}^{80} \geq 8$}
& $ N_{ { \bjet }} \geq 0$ & 35.7 & 5.6 & 33.8 & 8.3 & 70 & 10 & 73 \\ 
& $ N_{ { \bjet }} \geq 1$ & 31.6 & 5.7 & 24.8 & 6.4 & 56 & 10 & 53 \\ 
& $ N_{ { \bjet }} \geq 2$ & 15.5 & 3.8 & 11.6 & 3.3 & 27.1 & 6.0 & 29 \\ 
\midrule
\multirow{3}{*}{$N_{\mathrm{jet}}^{80} \geq 9$}
& $ N_{ { \bjet }} \geq 0$ & 4.3 & 1.3 & 4.2 & 1.8 & 8.5 & 2.0 & 8 \\ 
& $ N_{ { \bjet }} \geq 1$ & 4.5 & 1.3 & 2.9 & 1.5 & 7.4 & 1.8 & 7 \\ 
& $ N_{ { \bjet }} \geq 2$ & 2.34 & 0.95 & 1.69 & 0.89 & 4.0 & 1.2 & 6 \\ 
\midrule
\midrule
\multirow{2}{*}{$N_{\mathrm{jet}}^{50} \geq 8$}
& $\mjsigma \geq 340$ \GeV & 306 & 54 & 220 & 55 & 526 & 72 & 471 \\ 
& $\mjsigma \geq 500$ \GeV & 118 & 18 & 69 & 20 & 187 & 24 & 161 \\ 
\midrule
\multirow{2}{*}{$N_{\mathrm{jet}}^{50} \geq 9$}
& $\mjsigma \geq 340$ \GeV & 73 & 15 & 56 & 15 & 129 & 23 & 104 \\ 
& $\mjsigma \geq 500$ \GeV & 36.5 & 6.3 & 23.3 & 7.0 & 60 & 10 & 38 \\ 
\midrule
\multirow{2}{*}{$N_{\mathrm{jet}}^{50} \geq 10$}
& $\mjsigma \geq 340$ \GeV & 14.6 & 3.8 & 13.2 & 3.5 & 27.9 & 5.7 & 18 \\ 
& $\mjsigma \geq 500$ \GeV & 9.8 & 2.6 & 6.2 & 3.3 & 16.0 & 4.7 & 10 \\ 
\bottomrule
\end{tabular}
\caption{
  The expected SM background (and separately the multijet and leptonic
  contributions) and the observed number of data events. The SM background
  normalisations are obtained from fits to the data in control regions, as
  described in the text. 
}
\label{tab:fitResults}
\end{table}

\begin{figure}
\centering
\includegraphics[width=0.7\linewidth]{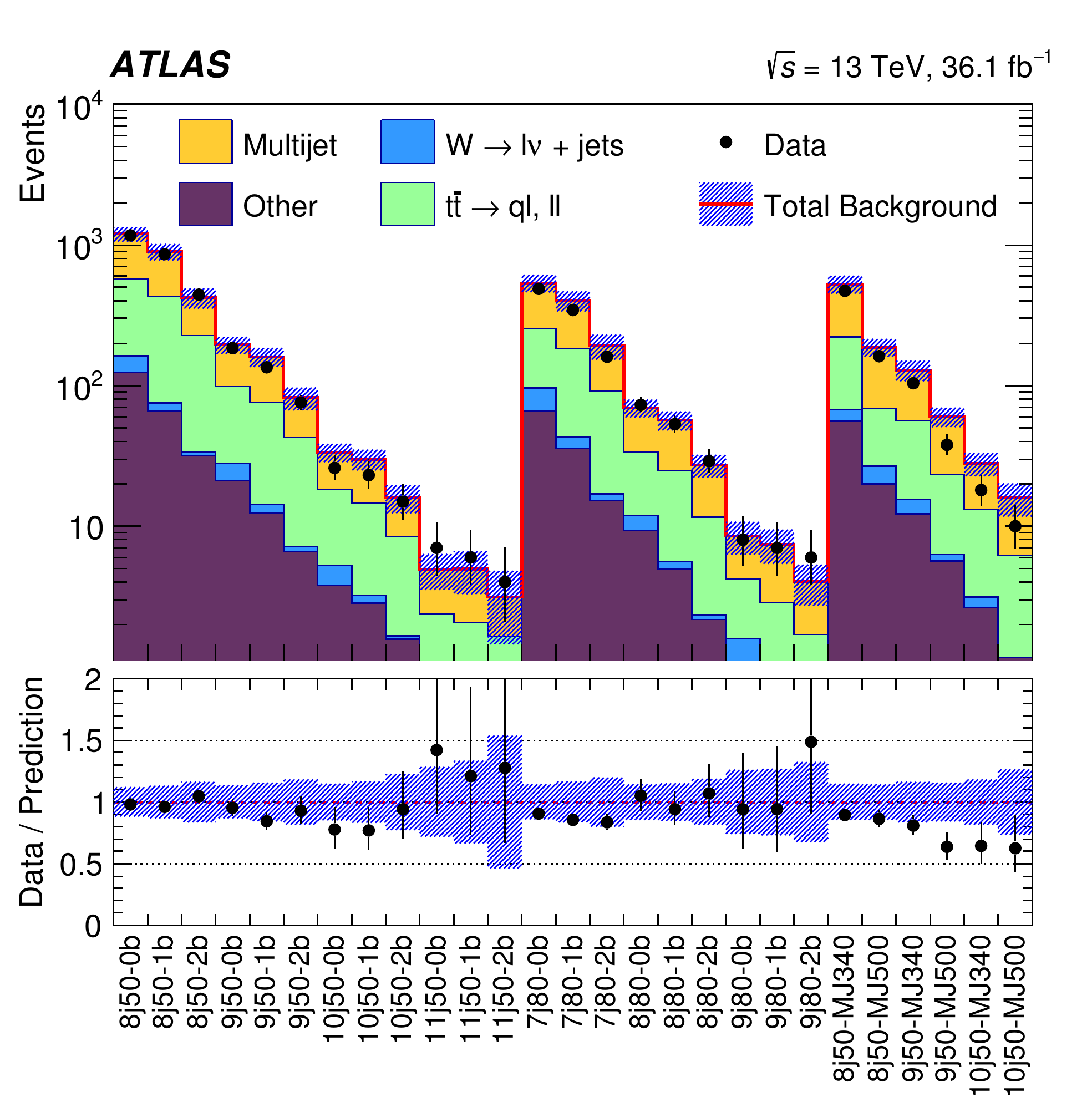}
\caption{Summary plot showing the data and SM predictions constrained by the likelihood fit for all signal regions.
Systematic and statistical uncertainties are shown in the blue hatched band, accounting for (anti-)correlations in their effects on different background components.~\panelnoredarrow
}
\label{fig:regionSummarySR_postFit}
\end{figure}

The full distributions of \METSig are shown for two of the most sensitive signal regions in Figure~\ref{fig:dataMC_SR_summary}.
For all signal regions, the data agree with the predicted \METSig distributions within the systematic uncertainties.

\begin{figure}[htb]
  \centering
  \begin{tabular}{cc}
    \subfloat[$\njfifty\geq11$, $\nbjet\geq0$]{\includegraphics[width=0.45\textwidth]{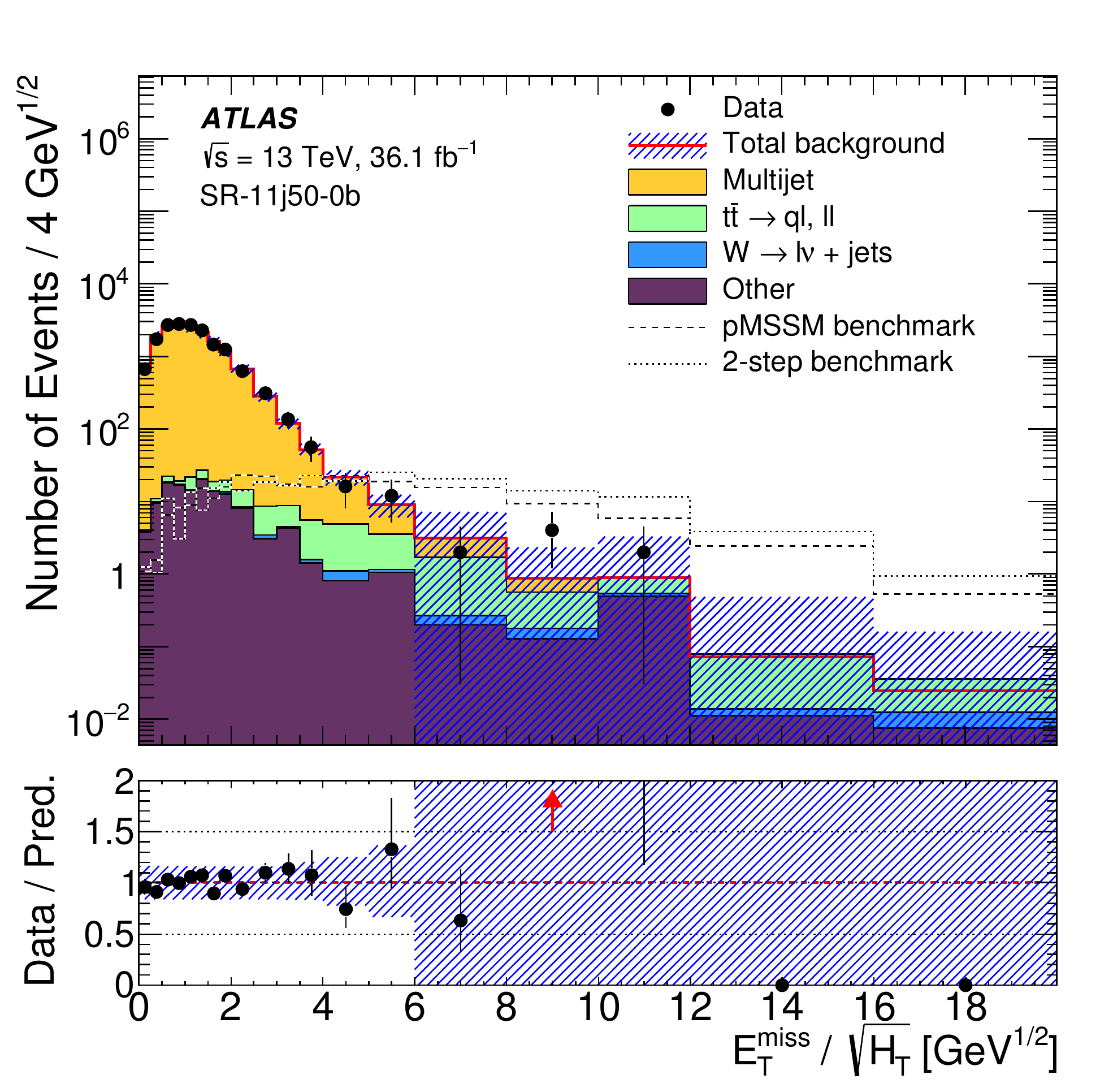}}\hspace{4ex}&
  \subfloat[$\njfifty\geq10$, $\mjsigma > 500~\GeV$]{\includegraphics[width=0.45\textwidth]{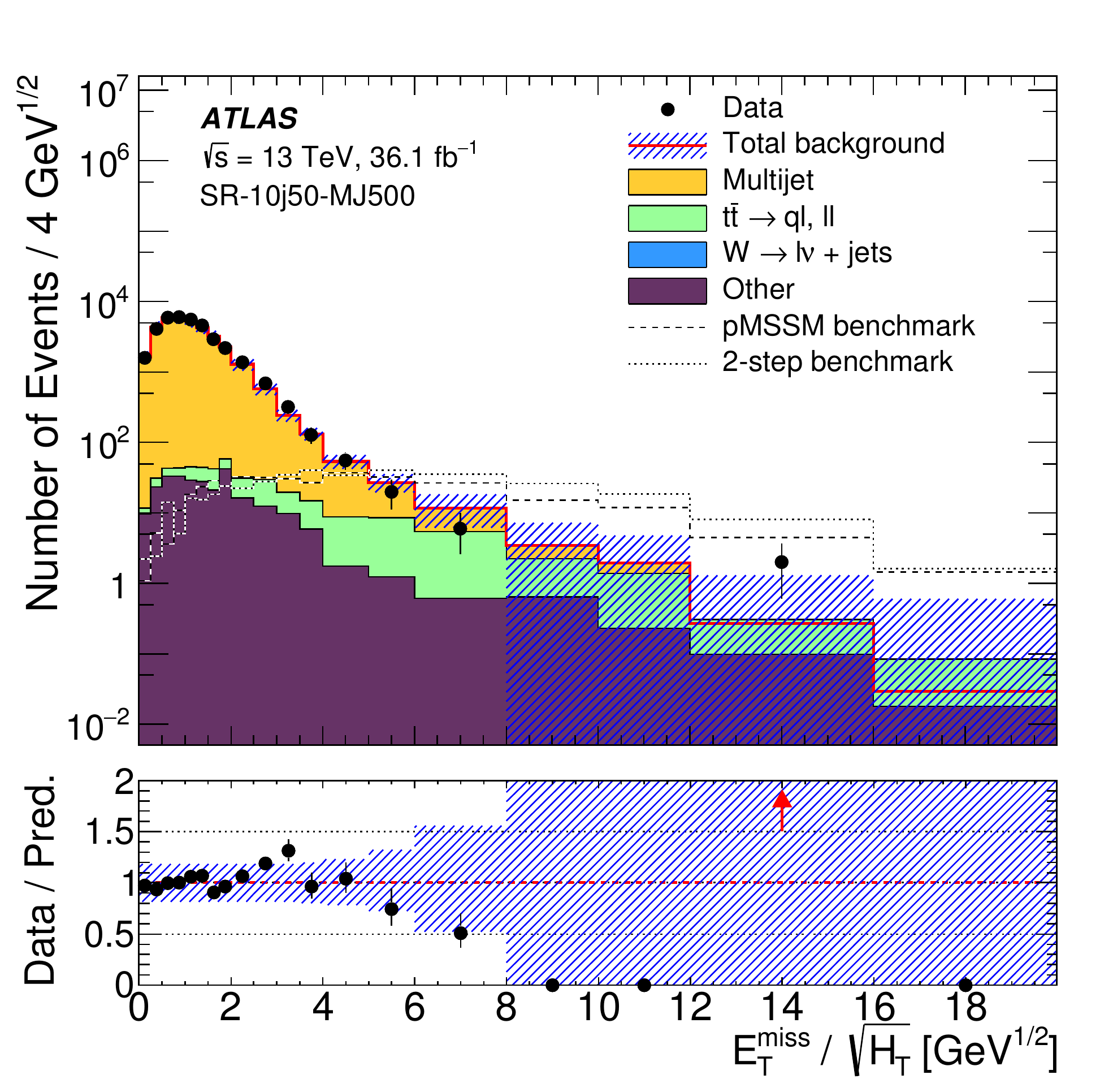}}\\
  \end{tabular}
\caption{Distributions of the \metsig for events in the 11-jet SR for the 50\,\GeV\ flavour channel, inclusive in \nbjet (a) and the 10-jet SR for the jet mass channel (b), with $\mjsigma>500~\GeV$. 
\scaledBackgrounds~\uncertaintyBand~\benchmarkModels~\panelnoredarrow~\redarrow}
\label{fig:dataMC_SR_summary}
\end{figure}

\clearpage

\Tab{\ref{tab:upperLimits}} quantifies the results of the fit to all signal regions.
When testing for a positive signal, the smallest $p_0$ value observed is 0.2, for $\njeighty\geq9$ and $\nbjet\geq2$.
The strongest limits set on the visible cross-section are of about 0.19~fb, for $\njfifty\geq11$ and $\nbjet\geq2$. 

\begin{table}
\begin{center}\renewcommand\arraystretch{1.4}
\begin{tabular*}{0.9\textwidth}{@{}@{\extracolsep{\fill}}llccccc}
\toprule
\multicolumn{2}{l}{Signal Region}  & $\langle \epsilon \sigma \rangle_{\mathrm{obs}}^{95}$ [fb] & $S_{\mathrm{obs}}^{95}$ & $S_{\mathrm{exp}}^{95}$ & $1-{\mathrm{CL}}_{\mathrm{b}}$ & $p_0$\\
    \midrule
\multirow{3}{*}{$N_{\mathrm{jet}}^{50} \geq 8$}
  & $ N_{ { \bjet }} \geq 0$ & 7.2 & 260 & $ { 270 }^{ +90 }_{ -70 }$ & 0.44 & 0.50\\ 
  & $ N_{ { \bjet }} \geq 1$ & 6.4 & 230 & $ { 250 }^{ +80 }_{ -60 }$ & 0.40 & 0.50\\ 
  & $ N_{ { \bjet }} \geq 2$ & 4.6 & 170 & $ { 160 }^{ +50 }_{ -40 }$ & 0.59 & 0.40\\ 
\midrule
\multirow{3}{*}{$N_{\mathrm{jet}}^{50} \geq 9$}
  & $ N_{ { \bjet }} \geq 0$ & 1.5 & 53 & $ { 58 }^{ +20 }_{ -15 }$ & 0.38 & 0.50\\ 
  & $ N_{ { \bjet }} \geq 1$ & 1.2 & 44 & $ { 55 }^{ +18 }_{ -14 }$ & 0.24 & 0.50\\ 
  & $ N_{ { \bjet }} \geq 2$ & 1.0 & 35 & $ { 38 }^{ +12 }_{ -9 }$ & 0.40 & 0.50\\ 
\midrule
\multirow{3}{*}{$N_{\mathrm{jet}}^{50} \geq 10$}
  & $ N_{ { \bjet }} \geq 0$ & 0.30 & 11 & $ { 15 }^{ +6 }_{ -4 }$ & 0.17 & 0.50\\ 
  & $ N_{ { \bjet }} \geq 1$ & 0.31 & 11 & $ { 15 }^{ +6 }_{ -4 }$ & 0.20 & 0.50\\ 
  & $ N_{ { \bjet }} \geq 2$ & 0.31 & 11 & $ { 12 }^{ +5 }_{ -3 }$ & 0.44 & 0.50\\ 
\midrule
\multirow{3}{*}{$N_{\mathrm{jet}}^{50} \geq 11$}
  & $ N_{ { \bjet }} \geq 0$ & 0.23 & 8.5 & $ { 6.3 }^{ +3.0 }_{ -1.5 }$ & 0.80 & 0.21\\ 
  & $ N_{ { \bjet }} \geq 1$ & 0.21 & 7.4 & $ { 6.5 }^{ +2.6 }_{ -1.7 }$ & 0.68 & 0.34\\ 
  & $ N_{ { \bjet }} \geq 2$ & 0.19 & 6.9 & $ { 6.0 }^{ +2.2 }_{ -1.3 }$ & 0.69 & 0.35\\ 
\midrule
\midrule
\multirow{3}{*}{$N_{\mathrm{jet}}^{80} \geq 7$}
  & $ N_{ { \bjet }} \geq 0$ & 3.1 & 110 & $ { 130 }^{ +40 }_{ -30 }$ & 0.27 & 0.50\\ 
  & $ N_{ { \bjet }} \geq 1$ & 2.7 & 100 & $ { 120 }^{ +40 }_{ -30 }$ & 0.23 & 0.50\\ 
  & $ N_{ { \bjet }} \geq 2$ & 1.7 & 60 & $ { 72 }^{ +22 }_{ -17 }$ & 0.26 & 0.50\\ 
\midrule
\multirow{3}{*}{$N_{\mathrm{jet}}^{80} \geq 8$}
  & $ N_{ { \bjet }} \geq 0$ & 0.80 & 29 & $ { 27 }^{ +10 }_{ -7 }$ & 0.60 & 0.40\\ 
  & $ N_{ { \bjet }} \geq 1$ & 0.62 & 22 & $ { 24 }^{ +9 }_{ -7 }$ & 0.40 & 0.50\\ 
  & $ N_{ { \bjet }} \geq 2$ & 0.49 & 18 & $ { 16 }^{ +6 }_{ -5 }$ & 0.59 & 0.41\\ 
\midrule
\multirow{3}{*}{$N_{\mathrm{jet}}^{80} \geq 9$}
  & $ N_{ { \bjet }} \geq 0$ & 0.22 & 7.8 & $ { 7.9 }^{ +3.4 }_{ -2.0 }$ & 0.47 & 0.50\\ 
  & $ N_{ { \bjet }} \geq 1$ & 0.21 & 7.5 & $ { 7.5 }^{ +2.8 }_{ -2.1 }$ & 0.48 & 0.50\\ 
  & $ N_{ { \bjet }} \geq 2$ & 0.22 & 8.0 & $ { 5.9 }^{ +2.6 }_{ -1.4 }$ & 0.81 & 0.20\\ 
\midrule
\midrule
\multirow{2}{*}{$N_{\mathrm{jet}}^{50} \geq 8$}
  & $\mjsigma \geq 340$ \GeV & 2.9 & 100 & $ { 130 }^{ +40 }_{ -30 }$ & 0.24 & 0.50\\ 
  & $\mjsigma \geq 500$ \GeV & 1.0 & 36 & $ { 48 }^{ +17 }_{ -13 }$ & 0.18 & 0.50\\ 
\midrule
\multirow{2}{*}{$N_{\mathrm{jet}}^{50} \geq 9$}
  & $\mjsigma \geq 340$ \GeV & 0.87 & 32 & $ { 42 }^{ +14 }_{ -11 }$ & 0.17 & 0.50\\ 
  & $\mjsigma \geq 500$ \GeV & 0.32 & 12 & $ { 20 }^{ +8 }_{ -6 }$ & 0.04 & 0.50\\ 
\midrule
\multirow{2}{*}{$N_{\mathrm{jet}}^{50} \geq 10$}
  & $\mjsigma \geq 340$ \GeV & 0.25 & 9.1 & $ { 14 }^{ +6 }_{ -4 }$ & 0.10 & 0.50\\ 
  & $\mjsigma \geq 500$ \GeV & 0.22 & 7.9 & $ { 11 }^{ +4 }_{ -3 }$ & 0.18 & 0.50\\ 
\bottomrule
\end{tabular*}
\end{center}
\caption{Left to right: 95\% CL upper limits on the visible cross section
($\langle\epsilon\sigma\rangle_{\mathrm {obs}}^{95}$) and on the number of signal
events ($S_{\mathrm {obs}}^{95}$ ).  The third column ($S_{\mathrm {exp}}^{95}$) shows the
95\% CL upper limit on the number of signal events, given the expected number
(and $\pm 1\sigma$ excursions on the expectation) of background events. 
The last two columns indicate $1-{\mathrm{CL}}_{\mathrm{b}}$, i.e. the complement of the $p$-value observed
for the background-only hypothesis, and the discovery $p$-value ($p_0$).
}
\label{tab:upperLimits}
\end{table}

\subsection{Exclusion limits}

Using the exclusion configuration defined in \Sect{\ref{sec:stats:hypotest}}, limits are set at the 95\% \mbox{CL} in the signal scenarios described in \Sect{\ref{sec:dataMC:signal}}.
Constraints from all 27 SRs are combined by considering only the SR with the best expected exclusion sensitivity at each signal model point.
These are illustrated in several parameter planes in Figures~\ref{fig:excl_pMSSM_2step} and \ref{fig:excl_gtt}.

In the $m_{\gluino}, m_{\chinoonepm}$ projection of the pMSSM, constraints are set such that $m_{\gluino}\gtrsim1600~\GeV$ is excluded for $m_{\chinoonepm}<600~\GeV$.
The limit falls to $m_{\gluino}\gtrsim1360~\GeV$ for $m_{\chinoonepm}\simeq800~\GeV$.

Limits are set up to $m_{\gluino} \approx 1800~\GeV$ for small LSP masses when considering the simplified model assuming a two-step cascade decay of the gluino.
For $m_{\gluino}\simeq800~\GeV$, models are excluded provided that \mbox{$m_{\ninoone}<475~\GeV$}.
The limits lie in the range $500 < m_{\ninoone} < 700~\GeV$ as the gluino mass increases to $m_{\gluino}=1600~\GeV$.

Simplified models of gluino-mediated top squark production are excluded for gluino masses up to $1500~\GeV$, as long as $m_{\ninoone}\lesssim600~\GeV$, when assuming that the squark is more massive than the gluino.
When RPC restrictions are relaxed, gluino masses between 625 and 1375~\GeV~are excluded for $400<m_{\ninoone}<800~\GeV$ in a scenario where the top squarks decay through an RPV coupling to $\bar{s}\bar{b}$.

\newcommand{\limitcontourdescr}{The solid maroon line indicates the observed limit, while the dashed blue line shows the expected limit. Experimental, MC theoretical and statistical uncertainties are shown in the yellow band. Dotted maroon lines delimit the variation of the observed limit within the $\pm1\sigma$ uncertainties in the signal cross-section at NLO+NLL accuracy.}

\begin{figure}
  \centering
  \begin{tabular}{c}
  \subfloat[pMSSM]{\includegraphics[width=0.7\linewidth]{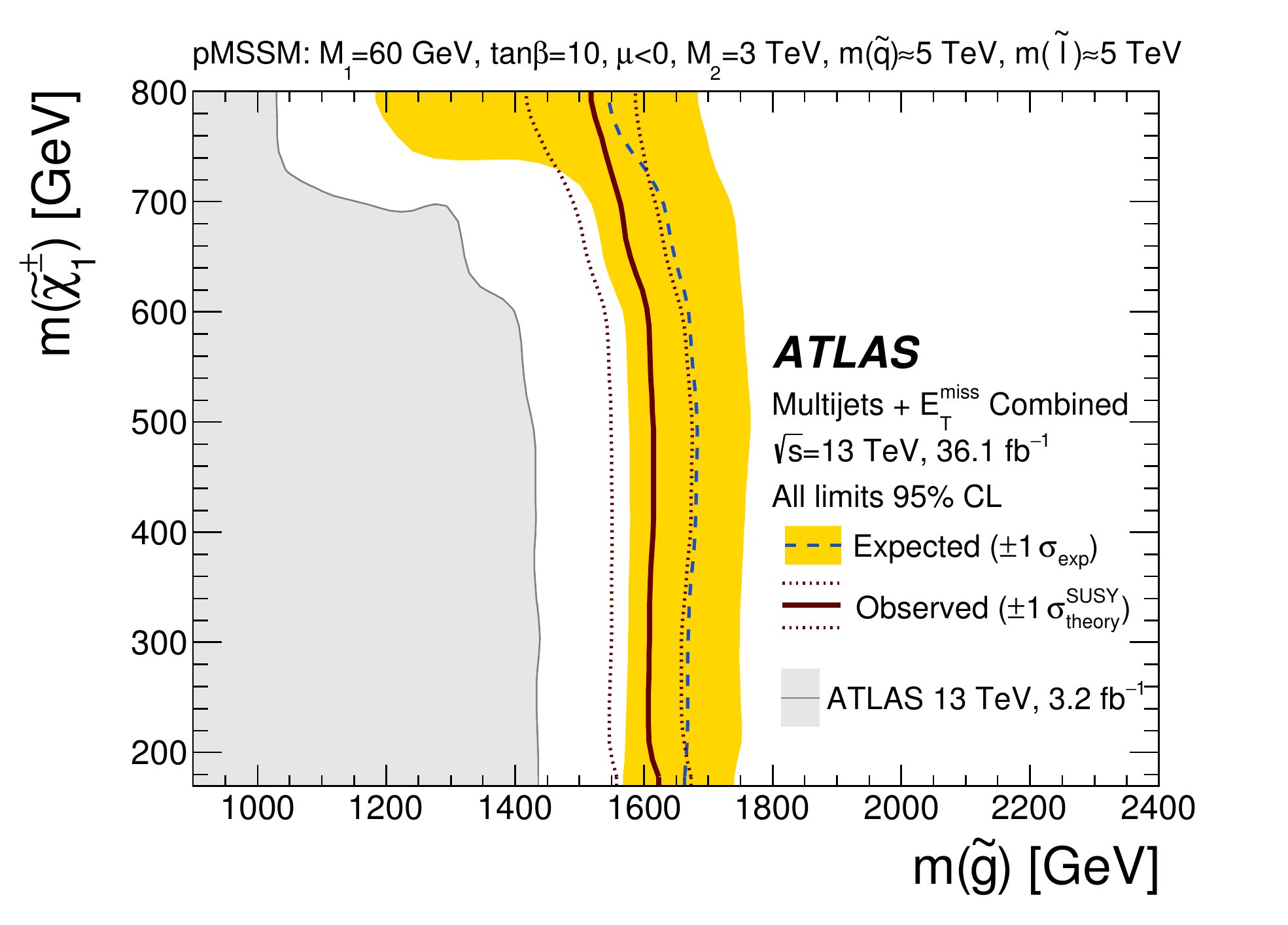}}\\
  \subfloat[Two-step decay]{\includegraphics[width=0.7\linewidth]{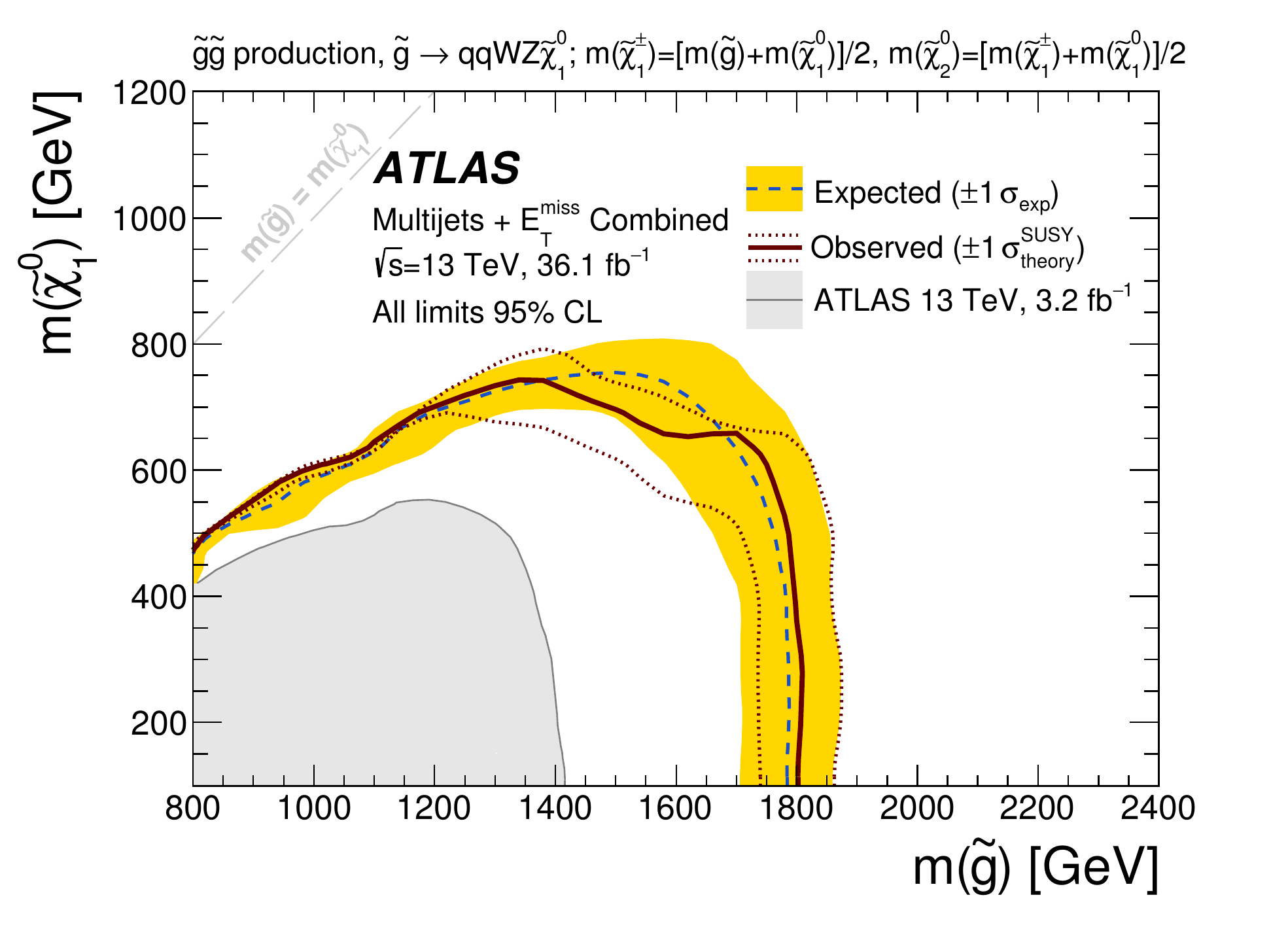}}
  \end{tabular}
  \caption{Exclusion contours in the $m_{\gluino}, m_{\chinoonepm}$ plane for the pMSSM (a) and the $m_{\gluino}, m_{\ninoone}$ plane in a simplified model with the gluino decaying via a two-step cascade (b).
  \limitcontourdescr{}
  }
  \label{fig:excl_pMSSM_2step}
\end{figure}

\begin{figure}
  \centering
  \begin{tabular}{c}
  \subfloat[Off-shell top squarks]{\includegraphics[width=0.7\linewidth]{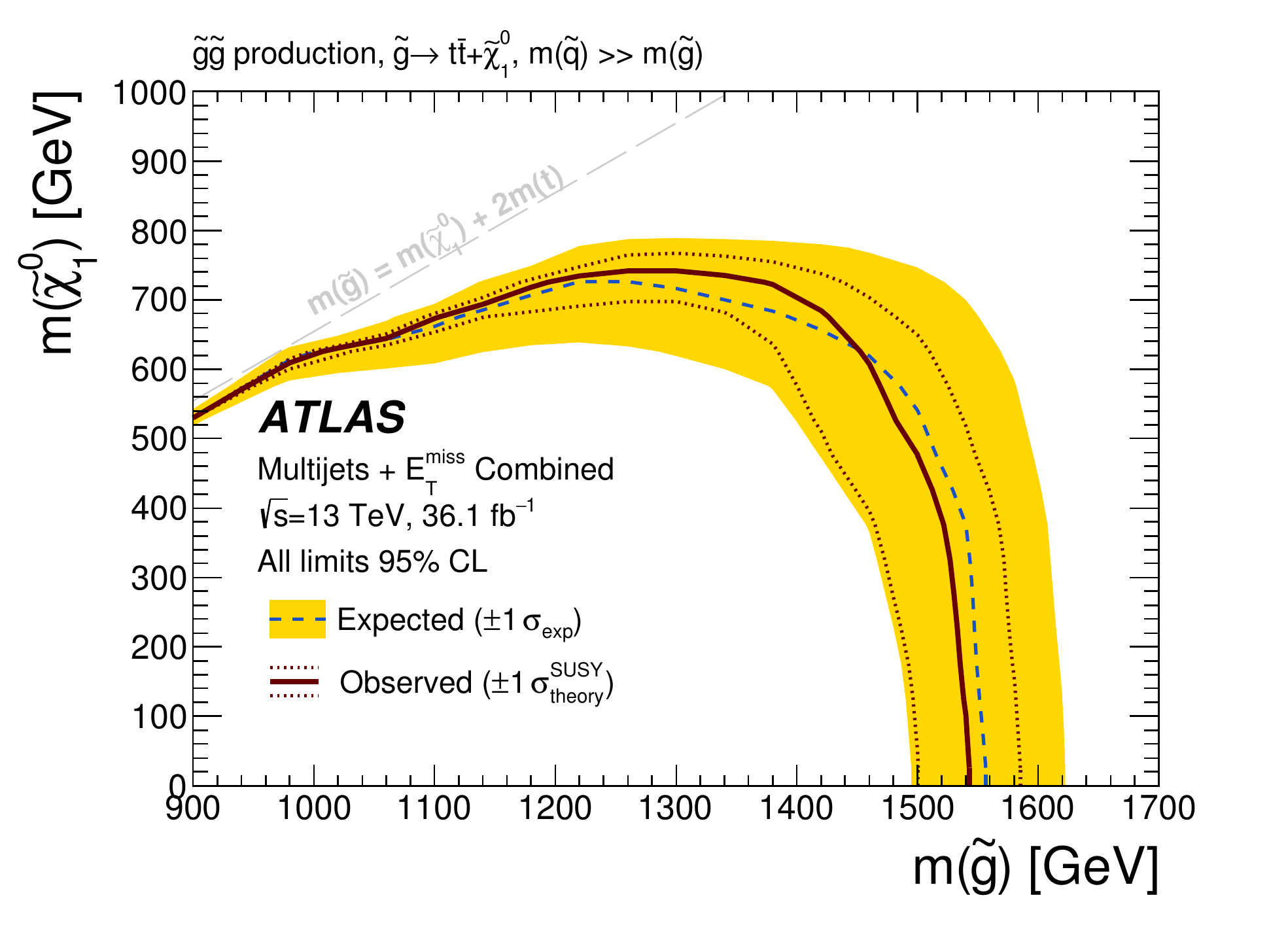}}\\
  \subfloat[RPV]{\includegraphics[width=0.7\linewidth]{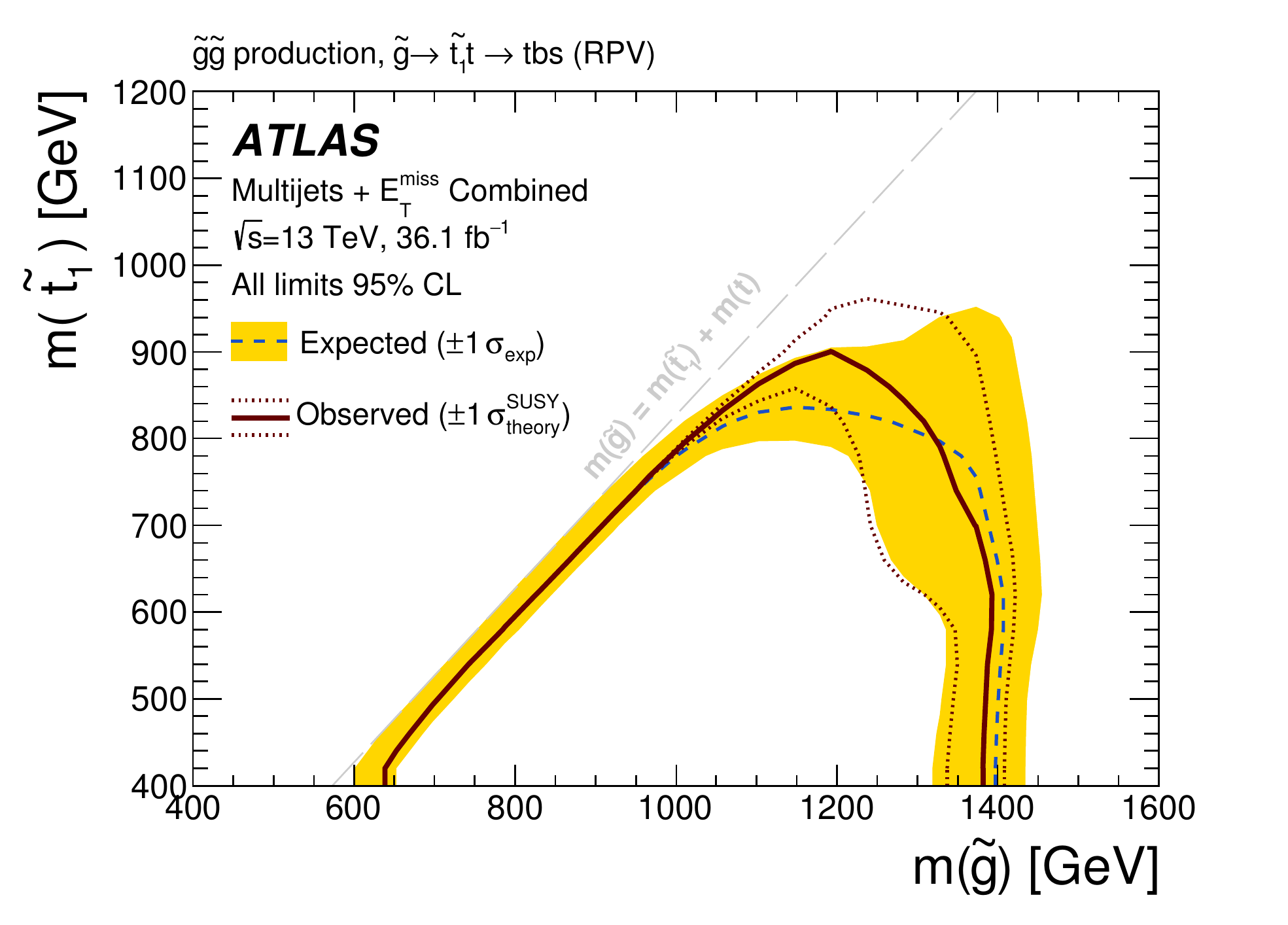}}
  \end{tabular}
  \caption{Exclusion contours in gluino-mediated top squark production scenarios, illustrated in the $m_{\gluino},\,m_{\ninoone}$ plane with off-shell squarks (a) and an $R$-parity-violating scenario plane
    in which the top squark decays via $\stopone\to sb$ (b), shown in the $m_{\gluino},\,m_{\stopone}$ plane.
  \limitcontourdescr{}
  }
  \label{fig:excl_gtt}
\end{figure}

\clearpage

\FloatBarrier

\section{Conclusion}
\label{sec:conclusion}

A search for heavy particles predicted by supersymmetric or other models of physics beyond the SM, that decay to produce large jet multiplicities in association with \met, was performed using \intlumi~\ifb\ of $\sqrt{s}=13$\,\TeV~LHC $pp$ collision data collected by ATLAS in 2015 and 2016.
No significant excesses over the Standard Model background were observed in signal regions selecting up to 11 jets with $\pT>50\,\GeV$ or 9 jets with $\pT>80\,\GeV$.
The largest-jet-multiplicity event observed in data had 13 jets with $\pT>~80\,\GeV$, while the greatest observed jet mass sum was $\mjsigma=1.3\,\TeV$.

Exclusion limits are placed on gluino production in supersymmetric signal scenarios with a range of model assumptions.
The tightest limits are set at $m_{\gluino}\approx1800$\,\GeV~in a simplified model assuming a two-step cascade decay via the \ninotwo and \chinoonepm.
A slice of the phenomenological MSSM is excluded for $m_{\gluino}<1360\,\GeV$, with tighter constraints at $m_{\gluino}\approx 1600\,\GeV$ for $m_{\chinoonepm}<600\,\GeV$.
When assuming that the gluino decays through off-shell top squarks, masses of the gluino below $1500\,\GeV$ are excluded provided $m_{\ninoone}<600\,\GeV$.
Limits are also set in an R-parity-violating model with baryon-number-violating couplings permitting
$\stopone\to\bar{s}\bar{b}$, such that the gluino mass must be greater than 1250--1375\,\GeV~when the top squark mass is in the range $400<m_{\stopone}<800\,\GeV$.

\section*{Acknowledgements}

% Acknowledgements for papers with collision data
% Version 06-Mar-2017

% Standard acknowledgements start here
%----------------------------------------------
We thank CERN for the very successful operation of the LHC, as well as the
support staff from our institutions without whom ATLAS could not be
operated efficiently.

We acknowledge the support of ANPCyT, Argentina; YerPhI, Armenia; ARC, Australia; BMWFW and FWF, Austria; ANAS, Azerbaijan; SSTC, Belarus; CNPq and FAPESP, Brazil; NSERC, NRC and CFI, Canada; CERN; CONICYT, Chile; CAS, MOST and NSFC, China; COLCIENCIAS, Colombia; MSMT CR, MPO CR and VSC CR, Czech Republic; DNRF and DNSRC, Denmark; IN2P3-CNRS, CEA-DSM/IRFU, France; SRNSF, Georgia; BMBF, HGF, and MPG, Germany; GSRT, Greece; RGC, Hong Kong SAR, China; ISF, I-CORE and Benoziyo Center, Israel; INFN, Italy; MEXT and JSPS, Japan; CNRST, Morocco; NWO, Netherlands; RCN, Norway; MNiSW and NCN, Poland; FCT, Portugal; MNE/IFA, Romania; MES of Russia and NRC KI, Russian Federation; JINR; MESTD, Serbia; MSSR, Slovakia; ARRS and MIZ\v{S}, Slovenia; DST/NRF, South Africa; MINECO, Spain; SRC and Wallenberg Foundation, Sweden; SERI, SNSF and Cantons of Bern and Geneva, Switzerland; MOST, Taiwan; TAEK, Turkey; STFC, United Kingdom; DOE and NSF, United States of America. In addition, individual groups and members have received support from BCKDF, the Canada Council, CANARIE, CRC, Compute Canada, FQRNT, and the Ontario Innovation Trust, Canada; EPLANET, ERC, ERDF, FP7, Horizon 2020 and Marie Sk{\l}odowska-Curie Actions, European Union; Investissements d'Avenir Labex and Idex, ANR, R{\'e}gion Auvergne and Fondation Partager le Savoir, France; DFG and AvH Foundation, Germany; Herakleitos, Thales and Aristeia programmes co-financed by EU-ESF and the Greek NSRF; BSF, GIF and Minerva, Israel; BRF, Norway; CERCA Programme Generalitat de Catalunya, Generalitat Valenciana, Spain; the Royal Society and Leverhulme Trust, United Kingdom.

The crucial computing support from all WLCG partners is acknowledged gratefully, in particular from CERN, the ATLAS Tier-1 facilities at TRIUMF (Canada), NDGF (Denmark, Norway, Sweden), CC-IN2P3 (France), KIT/GridKA (Germany), INFN-CNAF (Italy), NL-T1 (Netherlands), PIC (Spain), ASGC (Taiwan), RAL (UK) and BNL (USA), the Tier-2 facilities worldwide and large non-WLCG resource providers. Major contributors of computing resources are listed in Ref.~\cite{ATL-GEN-PUB-2016-002}.
%----------------------------------------------

\printbibliography

\newpage
% ATLAS Collaboration author list
% Data extracted on 19-Jun-2017 for paper reference SUSY-2016-13
\begin{flushleft}
{\Large The ATLAS Collaboration}

\bigskip

M.~Aaboud$^\textrm{\scriptsize 137d}$,
G.~Aad$^\textrm{\scriptsize 88}$,
B.~Abbott$^\textrm{\scriptsize 115}$,
O.~Abdinov$^\textrm{\scriptsize 12}$$^{,*}$,
B.~Abeloos$^\textrm{\scriptsize 119}$,
S.H.~Abidi$^\textrm{\scriptsize 161}$,
O.S.~AbouZeid$^\textrm{\scriptsize 139}$,
N.L.~Abraham$^\textrm{\scriptsize 151}$,
H.~Abramowicz$^\textrm{\scriptsize 155}$,
H.~Abreu$^\textrm{\scriptsize 154}$
% ...
\bigskip
\\
% ...
$^{12}$ Institute of Physics, Azerbaijan Academy of Sciences, Baku, Azerbaijan\\
% ...
$^{88}$ CPPM, Aix-Marseille Universit{\'e} and CNRS/IN2P3, Marseille, France\\
% ...
$^{115}$ Homer L. Dodge Department of Physics and Astronomy, University of Oklahoma, Norman OK, United States of America\\
% ...
$^{119}$ LAL, Univ. Paris-Sud, CNRS/IN2P3, Universit{\'e} Paris-Saclay, Orsay, France\\
% ...
$^{137}$ $^{(a)}$ Facult{\'e} des Sciences Ain Chock, R{\'e}seau Universitaire de Physique des Hautes Energies - Universit{\'e} Hassan II, Casablanca; $^{(b)}$ Centre National de l'Energie des Sciences Techniques Nucleaires, Rabat; $^{(c)}$ Facult{\'e} des Sciences Semlalia, Universit{\'e} Cadi Ayyad, LPHEA-Marrakech; $^{(d)}$ Facult{\'e} des Sciences, Universit{\'e} Mohamed Premier and LPTPM, Oujda; $^{(e)}$ Facult{\'e} des sciences, Universit{\'e} Mohammed V, Rabat, Morocco\\
% ...
$^{139}$ Santa Cruz Institute for Particle Physics, University of California Santa Cruz, Santa Cruz CA, United States of America\\
% ...
$^{151}$ Department of Physics and Astronomy, University of Sussex, Brighton, United Kingdom\\
% ...
$^{154}$ Department of Physics, Technion: Israel Institute of Technology, Haifa, Israel\\
$^{155}$ Raymond and Beverly Sackler School of Physics and Astronomy, Tel Aviv University, Tel Aviv, Israel\\
% ...
$^{161}$ Department of Physics, University of Toronto, Toronto ON, Canada\\
% ...
$^{a}$ Also at Department of Physics, King's College London, London, United Kingdom\\
$^{b}$ Also at Institute of Physics, Azerbaijan Academy of Sciences, Baku, Azerbaijan\\
$^{c}$ Also at Novosibirsk State University, Novosibirsk, Russia\\
$^{d}$ Also at TRIUMF, Vancouver BC, Canada\\
$^{e}$ Also at Department of Physics {\&} Astronomy, University of Louisville, Louisville, KY, United States of America\\
% ...
$^{*}$ Deceased
\end{flushleft}

% Created with xml2latex.py

\end{document}